# Climate and structure of the 8.2 ka event reconstructed from three speleothems from Germany

Sarah Waltgenbach[1], Denis Scholz[1], Christoph Spötl[2], Dana F. C. Riechelmann[1], Klaus P. Jochum[3], Jens Fohlmeister[4,5], Andrea Schröder-Ritzrau[6]

[1]Institut für Geowissenschaften, Johannes Gutenberg-Universität Mainz, J.-J.-Becher-Weg 21, 55128 Mainz, Germany

[2]Institut für Geologie, Universität Innsbruck, Innrain 52, 6020 Innsbruck, Austria

[3]Abteilung für Klimageochemie, Max-Planck-Institut für Chemie, Postfach 3060, 55020 Mainz, Germany

[4]Potsdam Institute for Climate Impact Research, Telegrafenberg, 14473 Potsdam, Germany

[5]Section 'Climate Dynamics and Landscape Development', GFZ German Research Centre for Geosciences, Telegrafenberg Building C, 14473 Potsdam, Germany

[6]Institut für Umweltphysik, Ruprecht-Karls Universität Heidelberg, Im Neuenheimer Feld 229, 69120 Heidelberg, Germany

*Corresponding author:

Sarah Waltgenbach

Institut für Geowissenschaften, Johannes Gutenberg-Universität Mainz, J.-J.-Becher-Weg 21, 55128 Mainz, Germany

Phone: +49 6131 39 25584     Fax: +49 6131 39 23070

Email: wenzs@uni-mainz.de

## Abstract

The most pronounced climate anomaly of the Holocene was the 8.2 ka cooling event. We present new $^{230}$Th/U-ages as well as high-resolution stable isotope and trace element data from three stalagmites from two different cave systems in Germany, which provide important information about the structure and climate variability of the 8.2 ka event in central Europe.

In all three speleothems, the 8.2 ka event is clearly recorded as a pronounced negative excursion of the $\delta^{18}$O values and can be divided into a 'whole event' and a 'central event'. All stalagmites show a similar structure of the event with a short negative excursion prior to the 'central event', which marks the beginning of the 'whole event'. The timing and duration of the 8.2.ka event are different for the individual records, which may, however, be related to dating uncertainties.

Whereas stalagmite Bu4 from Bunker Cave also shows a negative anomaly in the $\delta^{13}$C values and Mg content during the event, the two speleothems from the Herbstlabyrinth cave system do not show distinct peaks in the other proxies. This may suggest that the speleothem $\delta^{18}$O values recorded in the three stalagmites do not primarily reflect climate change at the cave site, but rather large-scale changes in the North Atlantic. This is supported by comparison with climate modelling data, which suggest that the negative peak in the speleothem $\delta^{18}$O values is mainly due to lower $\delta^{18}$O values of precipitation above the cave and that temperature only played a minor role. Alternatively, the other proxies may not be as sensitive as $\delta^{18}$O values to record this centennial-scale cooling event. This may particularly be the case for speleothem $\delta^{13}$C values as suggested by comparison with a climate modelling study simulating vegetation changes in Europe during the 8.2 ka event. Based on our records, it is not possible to resolve which of these hypotheses is most appropriate, but our multi-proxy dataset shows that regional climate evolution during the event was probably complex, although all $\delta^{18}$O records show a clear negative anomaly.

## 1. Introduction

In recent years, it became obvious that the Holocene, long considered as a period of relatively stable and warm climate, includes intervals of substantial climate variability (Bond et al., 1997; Mayewski et al., 2004; Wanner et al., 2011). The most pronounced climate anomaly was the 8.2 ka event, whose impact was widespread, including the North Atlantic, Greenland, Europe and the Middle East, parts of Africa, China, India as well as North America, Latin America at part of South America (Alley and Ágústsdóttir, 2005; Baldini et al., 2002). The catastrophic outburst of the ice-dammed meltwater lakes Agassiz and Ojibway in north-eastern Canada that resulted in a freshwater influx of more than $10^{14}$ m$^3$ into the North Atlantic Ocean is regarded as the most likely trigger of this distinct cooling event (Barber et al., 1999; Clarke et al., 2004; Rohling and Pälike, 2005; Thomas et al., 2007). This influx of cold meltwater led to a reduction of sea-surface salinity and sea-surface temperature of the western North Atlantic. The resulting reduced deep-water formation then led to perturbations of the thermohaline circulation and consequently to generally cooler conditions in the region of the North Atlantic (Alley and Ágústsdóttir, 2005; Mayewski et al., 2004; Morrill and Jacobsen, 2005). A decrease in solar activity associated with changes in ocean circulation (Bond et al., 2001), internal variability of the climate system (Renssen et al., 2007) as well as accelerated melting of the collapsing ice saddle that linked domes over Hudson Bay (Lochte et al., 2019; Matero et al., 2017) have also been considered as triggers of the event. Based on annual layer counting of ice cores, Thomas et al. (2007) constrained the length of the entire event to 160.5 ± 5.5 years (8.30-8.14 ka b2k), while the central event lasted 69 ± 2 years (8.26-8.19 ka b2k).

Since the 8.2 ka cooling event reflects the impact of a dramatic freshwater influx into the North Atlantic during an interglacial climate state, it can be considered as an analogue for future climate changes (Mayewski et al., 2004; Morrill and Jacobsen, 2005). In particular, the understanding of the impact of accelerated ice melting in a warmer world and the resulting hydrological changes in the mid and high latitudes, such as changes in the thermohaline circulation, may be improved substantially (Alley and Ágústsdóttir, 2005). Thus, a detailed investigation of the 8.2 ka event will contribute to the understanding of future climate anomalies.

Due to the short-lived nature of the event, a detailed investigation requires archives providing a high temporal resolution as well as an accurate and precise chronology. Ice cores, in which the 8.2 ka event was firstly recognized (Alley et al., 1997), and laminated lake sediments are important terrestrial archives. Although the 8.2 ka event has also been detected in pollen records (Ghilardi and O'Connell, 2013; Hede et al., 2010; Kofler et al., 2005; Seppä et al., 2005), the majority of terrestrial evidence for the event is based on $\delta^{18}O$ values (ice cores, speleothems, ostracods). For instance, the event was recorded in a benthic ostracod $\delta^{18}O$ record from Lake Ammersee in southern Germany with an estimated decrease of the average annual air temperature during the event of about -1.7 °C (von Grafenstein et al., 1998; 1999). Holmes et al. (2016) analysed $\delta^{18}O$ values of the fine fraction of sedimentary material from three lake sediment cores in western Ireland, and all three cores show an abrupt negative excursion, which lasted for about 200 years. They also used the ratio of *Betula* to *Corylus* pollen as a palynological marker for the 8.2 ka event (Holmes et al., 2016). In addition, the 8.2 ka event is clearly recorded in two speleothem $\delta^{18}O$ records from Katerloch Cave, Austria (Boch et al., 2009). A stalagmite from Père Noël Cave in southern Belgium shows a distinct shift in the $\delta^{18}O$ and $\delta^{13}C$ values as well as in Sr, Ba and Mg concentrations at 8.13 ± 0.03 ka BP (BP = AD 1950; Allan et al., 2018). Baldini et al. (2002) also identified an anomaly in Sr and P at 8.33 ± 0.08 ka BP in a stalagmite from Ireland, while a marked decrease in $\delta^{18}O$ values was later identified as an analytical artefact (Fairchild et al., 2006a). Recently Andersen et al. (2017) presented a $\delta^{18}O$ record from ostracods preserved in the varved lake sediments from Mondsee, Austria, and reported evidence of a 75-year-long interval of higher-than-average $\delta^{18}O$ values directly after the 8.2 ka event, possibly reflecting increased air temperatures in Central Europe.

Given their accurate and precise chronology in conjunction with high-resolution multi-proxy records (i.e., stable oxygen and carbon isotopes and several trace elements), speleothems are ideal archives to investigate short-lived climate events. It is crucial, however, to replicate proxy records. The aim of this study is to provide detailed insights into the structure of the 8.2 ka event in Europe based on speleothems. To this end, three stalagmites (Bu4, HLK2 and TV1) from two cave systems in Germany (Bunker Cave and Herbstlabyrinth) were investigated using high-resolution stable oxygen and carbon isotope as well as trace element data. The study sites are known to be sensitive to changes in precipitation and temperature in relation to the North Atlantic (Fohlmeister et al., 2012; Mischel et al., 2017).

## 2. Samples

The three stalagmites were previously already analysed at lower resolution, and the 8.2 ka event was clearly recorded as a distinct negative $\delta^{18}O$ excursion (Figure 1; Fohlmeister et al., 2012; Mischel et al., 2017). The resolution of the $\delta^{18}O$ record from Bu4 was about eight years, while that of the two records from the Herbstlabyrinth cave system was 43 (HLK2) and 67 years (TV1), respectively. In this study we analysed the 8.2 ka section of the three stalagmites at very high resolution (about 3.4 to 5.9 yr/sample).

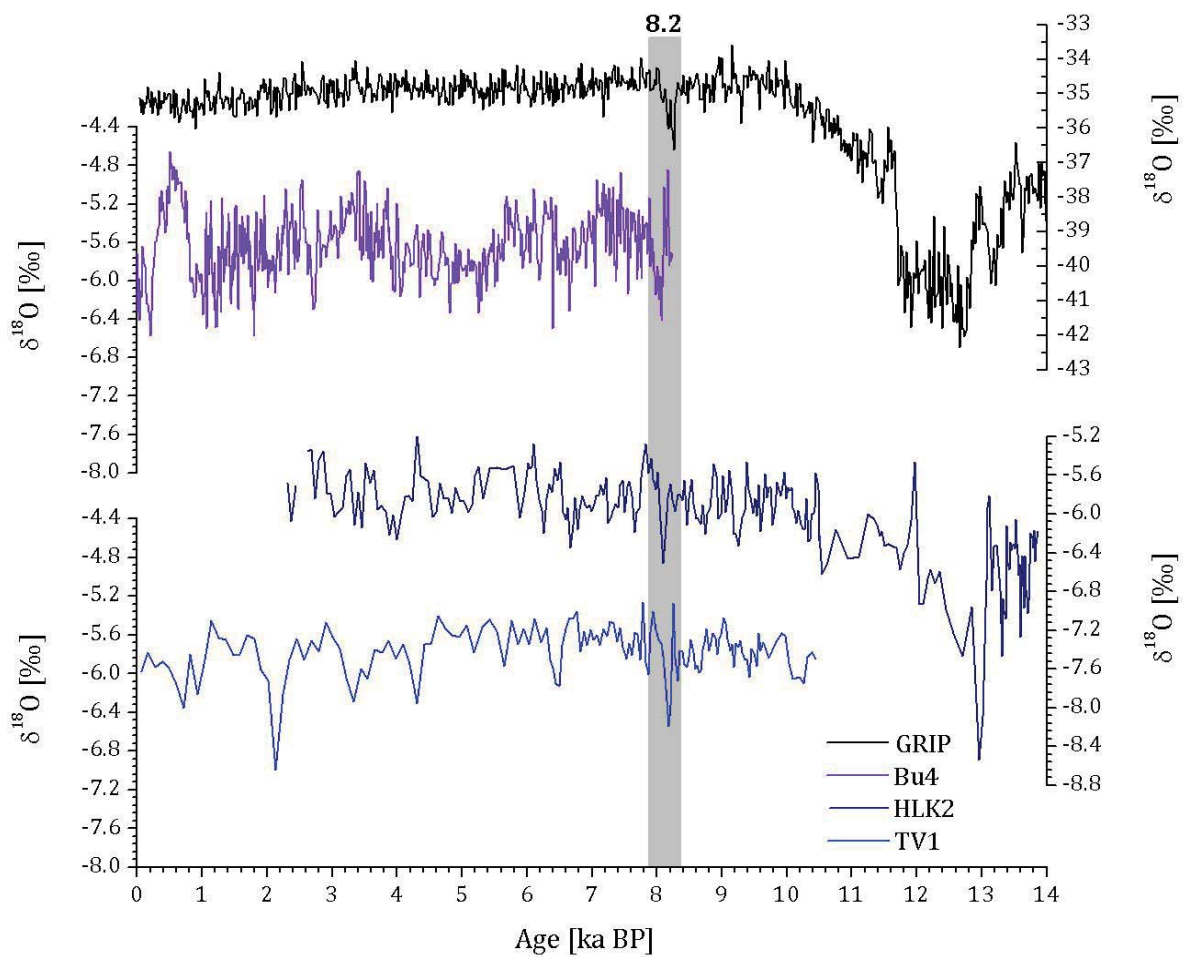

Figure 1: Compilation of the GRIP $\delta^{18}O_{ice}$ record (Rasmussen et al., 2006; Vinther et al., 2006) and the published low-resolution $\delta^{18}O$ records of the individual stalagmites (Bu4, HLK2, and TV1) from the two cave systems for the last 14

*ka. The grey bar highlights the 8.2 ka cooling event. All speleothem records are shown on the published age models, which were constructed with* StalAge *(Scholz and Hoffmann, 2011) for the stalagmites from the Herbstlabyrinth cave system and with* iscam *(Fohlmeister, 2012) for stalagmite Bu4 from Bunker Cave.*

## 2.1 Bunker Cave

Bunker Cave is located in western Germany near Iserlohn (51°22'03''N, 7°39'53''E, Figure 2) in the Rhenish Slate Mountains and is part of the 3.5 km long Bunker-Emst-Cave system (Riechelmann et al., 2011). The cave system developed in Middle to Upper Devonian limestone and was discovered in 1926 during road works (Fohlmeister et al., 2012; Riechelmann et al, 2011, 2012). The southern entrance of the cave system is located 184 m above sea level on a south-dipping hill slope (Riechelmann et al., 2011). The entrance to Emst-Cave is situated ca. 13 m above. The 15 to 30 m-thick bedrock is covered by up to 70 cm of brown and loamy soil as well as a vegetation, consisting entirely of C3 plants (i.e., mainly ash, beech and scrub vegetation; Fohlmeister et al., 2012; Grebe, 1993; Riechelmann et al., 2011). In total, six speleothems were removed from Bunker Cave (Bu1, Bu2, Bu3, Bu4, Bu5, and Bu6), but only stalagmite Bu4 grew during the 8.2 ka cooling event (Figure 3). From August 2006 to August 2013, a seven year-long monitoring programme was performed in and above Bunker Cave, which contributes to a better understanding and interpretation of the different proxy records (Immenhauser et al., 2010; Riechelmann et al., 2011, 2014, 2017; Wackerbarth et al., 2012). Furthermore, several speleothems from Bunker Cave were studied in terms of past climate variability (Fohlmeister et al., 2012; Weber et al., 2018).

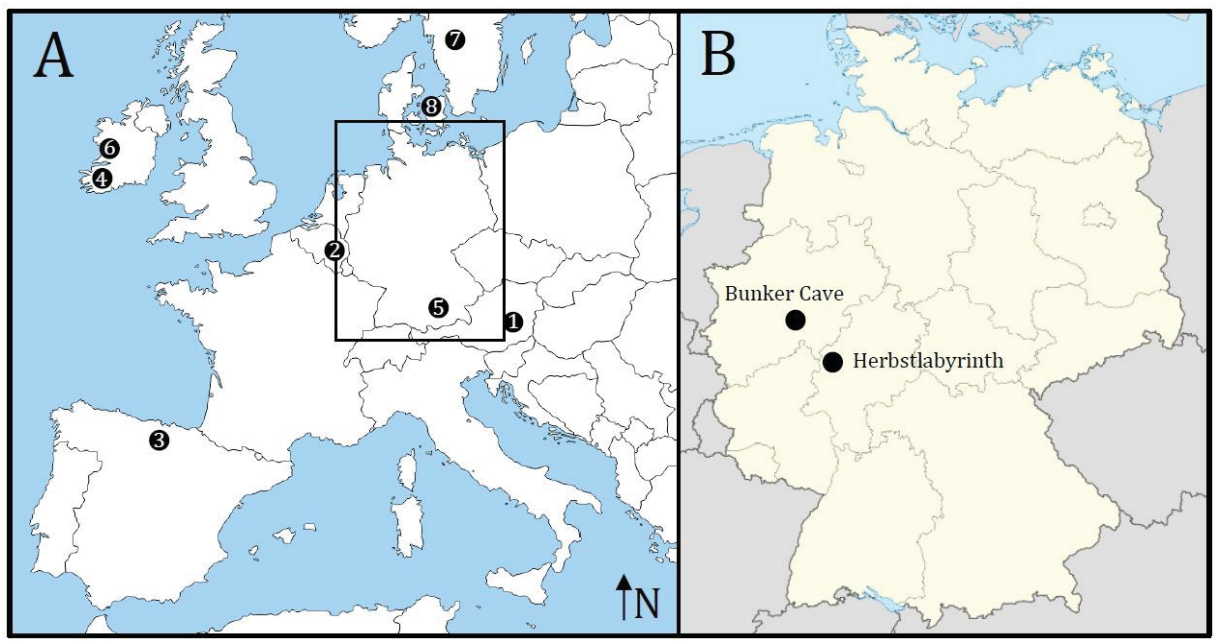

*Figure 2: A: Map of Europe showing the locations of study sites mentioned in section 1 as well as subsections 5.1 and 5.4: 1) Katerloch Cave (Boch et al., 2009), 2) Père Noël Cave (Allan et al., 2018), 3) Kaite Cave (Domínguez-Villar et al., 2012), 4) Crag Cave (Baldini et al., 2002), 5) Lake Ammersee (von Grafenstein et al., 1998, 1999), 6) Loch Avolla, Loch Gealáin and Lough Corrib (Holmes et al., 2016), 7) Lake Flarken (Seppä et al., 2005), 8) Lake Højby Sø (Hede et al., 2010). B: Map of Germany showing the locations of the two cave systems investigated in this study.*

## 2.2 Herbstlabyrinth cave system

The Herbstlabyrinth cave system is situated near Breitscheid in the Rhenish Slate Mountains in Central Germany (Figure 2). The ca. 9 km-long cave system developed in Devonian limestone and is located 435 m above sea level (Mischel et al., 2015, 2017). A ca. 60 cm-thick Cambisol (Terra fusca) is covered by patchy vegetation, consisting mainly of meadow and deciduous forest (Mischel et al., 2015). The cave system has four levels and shows a variety of different speleothems (Mischel et al., 2015). Since September 2010, a five year-long monitoring programme was performed in and above the cave system (Mischel et al., 2015). In total, three Holocene stalagmites (HLK2, NG01 and TV1) were removed from the Herbstlabyrinth cave system. Stalagmite NG01 shows a hiatus between 8.81 ± 0.01 ka BP and 7.65

± 0.06 ka BP, but stalagmites HLK2 and TV1 (Figure 3) grew during the 8.2 ka event and are included in this study. A detailed description of these two speleothems can be found in Mischel et al. (2017).

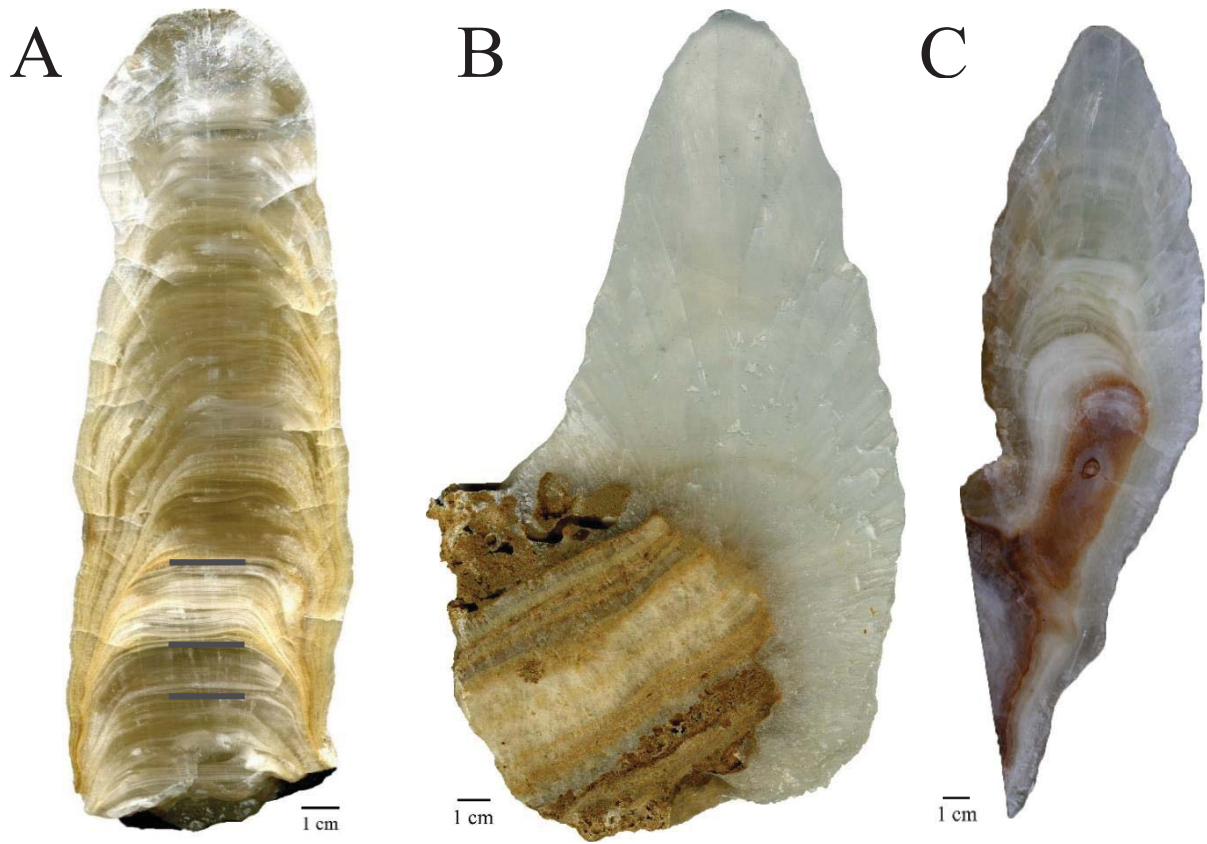

*Figure 3: Pictures of the three studied stalagmites: Bu4 (A), TV1 (B) and HLK2 (C). The lines on the picture of Bu4 indicate the positions of the three hiatuses (compare Fig. 4A).*

## 3. Methods

### 3.1 $^{230}$Th/U dating

$^{230}$Th/U-dating of the two stalagmites from the Herbstlabyrinth cave system was performed at the Max Planck Institute for Chemistry (MPIC), Mainz, with a multi collector inductively coupled plasma mass spectrometer (MC-ICP-MS, Nu Instruments). The results and further details about the dating of these stalagmites are reported in Mischel et al. (2017). Stalagmite Bu4 from Bunker Cave was analysed by thermal ionisation mass spectrometry (TIMS) at the Heidelberg Academy of Sciences (Fohlmeister et

al., 2012). Because of the relatively low uranium content of the speleothems from Bunker Cave, the age uncertainties are comparably large. Therefore, in the framework of this study, 17 additional $^{230}$Th/U-ages for stalagmite Bu4 were determined by MC-ICP-MS. For this purpose, several samples were drilled from the growth axis of the stalagmite using a hand-held dental drill (MICROMOT 50/E, Proxxon). Because of the relatively low U content of the speleothem, we used a relatively large sample mass of ca. 300 mg.

In a first step, the samples were dissolved in 7N HNO$_3$, and a mixed $^{229}$Th-$^{233}$U-$^{236}$U spike was added (Gibert et al., 2016). The samples were then dried down, and organic matter was removed by addition of concentrated HNO$_3$, HCl, and H$_2$O$_2$. After evaporation, the samples were dissolved in 6N HCl. These individual sample solutions were then passed through three Bio-Rad AG1-X8 columns to separate the Th and U fractions. Details about this ion exchange column chemistry are included in Hoffmann et al. (2007) and Yang et al (2015). For the Th and U-analyses by MC-ICP-MS, the separated Th and U fractions were dissolved in 2 ml 0.8N HNO$_3$. The Th and U samples were measured separately using a standard-sample bracketing procedure, in which each sample measurement is embraced by standard measurements (Hoffmann et al., 2007). The standard used for uranium measurements is the reference material CRM 112-A (New Brunswick Laboratory), the thorium standard is an in-house standard with previously calibrated ratios of $^{232}$Th, $^{230}$Th, and $^{229}$Th. Analytical details are described in Obert et al. (2016). The $^{230}$Th/U-ages are given in BP and for the calculation of individual age-models of all three stalagmites, we used the algorithm *StalAge* (Scholz and Hoffmann, 2011).

### 3.2 δ$^{18}$O and δ$^{13}$C values

For stable isotope measurements, the stalagmites were sampled along the growth axis using a computer-controlled MicroMill (Merchantek, New Wave). Depending on the growth rate of each speleothem, the spatial resolution was either 50 μm (HLK2) or 100 μm (Bu4 and TV1). Thus, all three stalagmites have a temporal resolution between 3 and 5 years. All stable isotope measurements were carried out with a ThermoFisher Delta$^{plus}$XL isotope ratio mass spectrometer linked to a Gasbench II at the University of Innsbruck with a 1σ-precision of 0.08 ‰ for δ$^{18}$O and 0.06 ‰ for δ$^{13}$C (Spötl, 2011). All stable isotope values are reported relative to the Vienna PeeDee Belemnite standard (VPDB).

### 3.3 Trace elements

Trace element measurements (Ca, Mg, P, Sr, Ba, and U) were performed at the MPIC, Mainz, by laser-ablation inductively coupled plasma mass spectrometry (LA-ICPMS). A Nd:YAG UP-213 nm laser ablation system from New Wave Research was coupled to a Thermo Scientific ELEMENT 2 single-collector sector-field mass spectrometer (Jochum et al., 2007, 2012). The reference materials used for the trace element measurements were the synthetic reference glass NIST SRM 612 (Jochum et al., 2005) as well as the carbonate reference material MACS-3 (http://georem.mpch-mainz.gwdg.de/; 28.05.2019). Calcium was used as an internal standard. The line-scan measurements were performed using a spot size of 100 µm, a repetition rate of 10 Hz and a scan speed of 10 µm s$^{-1}$.

### 4. Results

#### 4.1 $^{230}$Th/U-dating

The results of $^{230}$Th/U-dating of speleothem Bu4 are presented in Table 1. The $^{238}$U-content varies between 0.047 (Bu4-5.2) and 0.14 µg g$^{-1}$ (Bu4-5.1). The $^{232}$Th-content of the stalagmite is <1 ng g$^{-1}$ (Table 1). Only two samples (Bu4-4.3 and Bu4-2.6) show a $^{232}$Th-content of more than 1 ng g$^{-1}$. The ($^{230}$Th/$^{232}$Th) ratio of all samples varies between 2 and 260 (Table 1). The ($^{230}$Th/$^{232}$Th) ratio is an indicator of the degree of detrital contamination and for ($^{230}$Th/$^{232}$Th) > 20, the contamination can be considered negligible. For six of the 17 samples, ($^{230}$Th/$^{232}$Th) < 20 necessitating a correction for detrital contamination (Schwarcz, 1989). The $^{230}$Th/U-ages of stalagmite Bu4 are between 0.12 ± 0.11 and 8.08 ± 0.17 ka BP and show 2σ-age uncertainties between 49 (Bu4-5.1) and 210 years (Bu4-2.6). Radiocarbon analyses from the top section of speleothem Bu4 suggest that the stalagmite was actively growing when removed (Fohlmeister et al., 2012; Welte et al., 2016). Bu4 shows evidence for three potential hiatuses (Fig. 3A): Based on thin sections, Fohlmeister et al. (2012) described a layer of coralloid calcite at 15 cm distance from top, which probably reflects a growth stop of the stalagmite because coralloids form from aerosols. However, this hiatus is relatively short and was not resolved by the published $^{230}$Th/U-ages (Fohlmeister et al., 2012). A similar coralloid layer occurs at 17 cm distance from top (Figs. 3A and 4). Finally, another potential hiatus occurs at 18.5 cm distance from top that is also clearly visible in the new trace element data, which show distinct positive peaks in Al, Fe, and Si at 18.5 cm distance from top (supplemental Figure A.1). This growth stop corresponds to a section with dendritic crystals and a detrital layer, which may also explain the distinct maxima in several trace elements and occurred after the 8.2 ka event, which is defined between 19.2 and 20 cm distance from top by the $\delta^{18}$O record.

The new $^{230}$Th/U-ages clearly reveal the hiatuses at 15 and 18.5 cm distance from top and also strongly indicate the hiatus at 17 cm distance from top. The new age model of stalagmite Bu4 was calculated using StalAge (Scholz and Hoffmann, 2011) and includes 17 new MC-ICP-MS $^{230}$Th/U-ages as well as 11 previous TIMS $^{230}$Th/U-ages from Fohlmeister et al. (2012). To account for hiatuses, the original publication presenting StalAge (Scholz and Hoffmann, 2011) suggests to divide up the age model and individually fit the corresponding sections. One of the basic assumptions of StalAge is that it always uses sets of three or more data points for piecewise construction of the age model. Thus, calculating individual age models for the individual sections between the hiatuses is not straightforward for Bu4, because only two ages are available for some of these sections (Fig. 4). This problem can be circumvented by inserting virtual ages at the corresponding sections without adding chronological information (i.e., with very large uncertainties). The same procedure has also been used for other records (e.g., Wassenburg et al., 2016). The new age model is shown in Fig. 4 and clearly reveals the three hiatuses, in particular the one at 18.5 cm distance from top. We note that the new age model shows an age inversion at the hiatus at 17 cm distance from top, even if not significant within the very large error. This could only be improved by additional dating around the hiatus at 17 cm distance from top. However, since the focus of this study is on the 8.2 ka event, which is recorded in the bottom section of Bu4, this is not crucial for the results presented in this paper. The age model for the bottom section is well constrained by five $^{230}$Th/U-ages.

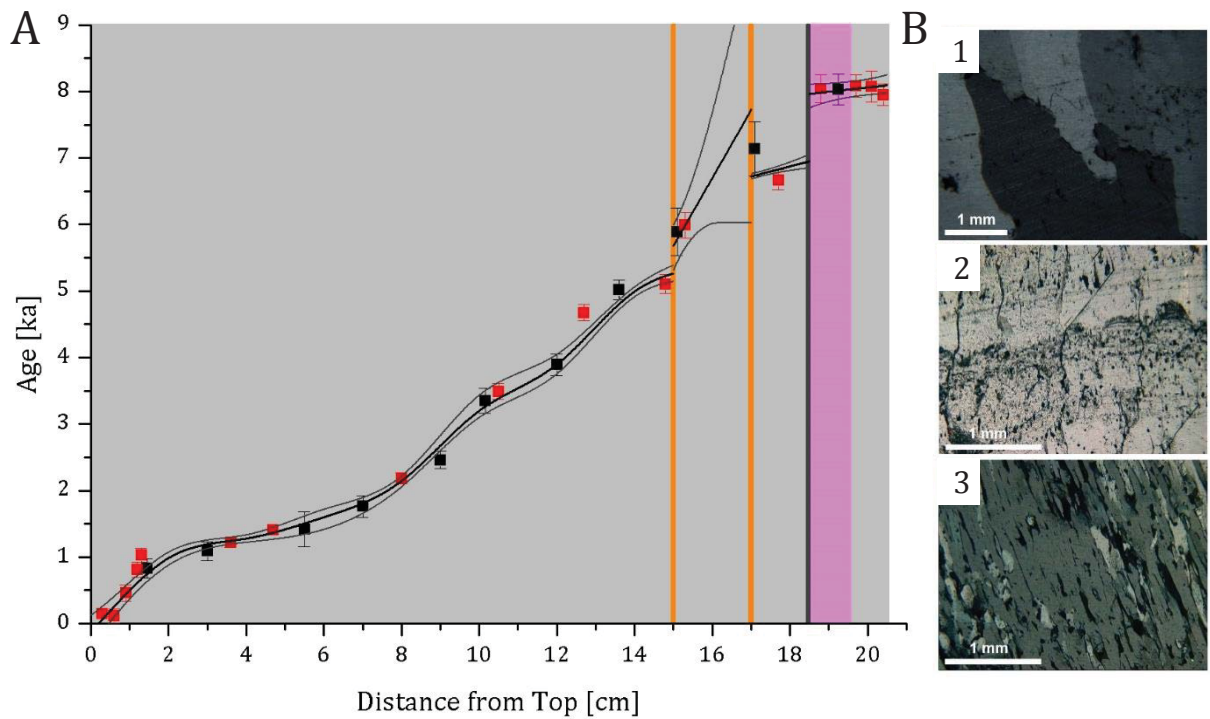

Figure 4: A: Age model for stalagmite Bu4. $^{230}$Th/U-ages marked with black squares indicate the ages measured with TIMS (Fohlmeister et al., 2012), and the red squares indicate the new MC-ICP-MS ages. The light grey colour marks columnar fabric, orange bars show the two coralloid layers, which are indicative of short hiatuses, and the pink colour indicates dendritic fabric. The dark grey bar indicates the growth stop of the speleothem, which is also visible in the trace element data. B: 1) columnar fabric under crossed nicols, 2) coralloid layer, 3) dendritic fabric under crossed nicols.

Table 1: U and Th activity ratios and [230Th/U]-ages for stalagmite Bu4 from Bunker Cave measured by MC-ICP-MS. All errors are shown at the 2σ level and activity ratios are indicated by parentheses.

| Sample ID | Depth [cm] | $^{238}$U [μg g$^{-1}$] | $^{232}$Th [ng g$^{-1}$] | ($^{234}$U/$^{238}$U) | ($^{230}$Th/$^{238}$U) | ($^{230}$Th/$^{232}$Th) | ($^{234}$U/$^{238}$U)$_{initial}$ | age$_{uncorrected}$ [ka BP] | age$_{corrected}$[a] [ka BP] |
|---|---|---|---|---|---|---|---|---|---|
| Bu4-5.1 | 0.3 | 0.1345 ± 0.0011 | 0.539 ± 0.012 | 1.5122 ± 0.0078 | 0.00210 ± 0.00067 | 2.406 ± 0.32 | 1.5124 ± 0.0077 | 0.228 ± 0.030 | 0.152 ± 0.049 |
| Bu4-4.1 | 0.6 | 0.05604 ± 0.00032 | 0.1982 ± 0.0045 | 1.5155 ± 0.0031 | 0.0016 ± 0.0016 | 2.2 ± 1.3 | 1.5157 ± 0.0030 | 0.18 ± 0.11 | 0.12 ± 0.11 |
| Bu4-5.2 | 0.9 | 0.04682 ± 0.00037 | 0.4613 ± 0.0047 | 1.5032 ± 0.0083 | 0.0063 ± 0.0017 | 2.752 ± 0.32 | 1.5039 ± 0.0081 | 0.647 ± 0.076 | 0.46 ± 0.12 |
| Bu4-5.3 | 1.2 | 0.06593 ± 0.00047 | 0.0563 ± 0.0014 | 1.4807 ± 0.0065 | 0.0110 ± 0.0015 | 40.1 ± 5.5 | 1.4818 ± 0.0065 | 0.83 ± 0.11 | 0.81 ± 0.11 |
| Bu4-4.2 | 1.3 | 0.06711 ± 0.00039 | 0.2719 ± 0.0065 | 1.5108 ± 0.0027 | 0.0142 ± 0.0015 | 11.5 ± 1.1 | 1.5123 ± 0.0027 | 1.11 ± 0.10 | 1.03 ± 0.11 |
| Bu4-1.1 | 3.6 | 0.07528 ± 0.00061 | 0.1948 ± 0.0036 | 1.4985 ± 0.0085 | 0.0167 ± 0.0010 | 20.5 ± 1.2 | 1.5002 ± 0.0086 | 1.274 ± 0.070 | 1.224 ± 0.073 |
| Bu4-2.1 | 4.7 | 0.07130 ± 0.00044 | 0.1339 ± 0.0047 | 1.5027 ± 0.0030 | 0.01930 ± 0.00089 | 32.2 ± 1.8 | 1.5047 ± 0.0029 | 1.447 ± 0.064 | 1.411 ± 0.065 |
| Bu4-2.2 | 8.0 | 0.07597 ± 0.00048 | 0.1848 ± 0.0065 | 1.5680 ± 0.0027 | 0.0312 ± 0.0012 | 39.9 ± 2.0 | 1.5715 ± 0.0028 | 2.233 ± 0.079 | 2.189 ± 0.081 |
| Bu4-2.3 | 10.5 | 0.06707 ± 0.00042 | 0.0389 ± 0.0017 | 1.5499 ± 0.0035 | 0.0488 ± 0.0017 | 258.1 ± 14.4 | 1.5553 ± 0.0035 | 3.50 ± 0.12 | 3.49 ± 0.12 |
| Bu4-2.4 | 12.7 | 0.07810 ± 0.00051 | 0.3548 ± 0.0041 | 1.6115 ± 0.0029 | 0.0677 ± 0.0017 | 46.3 ± 1.2 | 1.6196 ± 0.0029 | 4.75 ± 0.11 | 4.67 ± 0.12 |
| Bu4-2.5 | 14.8 | 0.07767 ± 0.00043 | 0.7955 ± 0.0083 | 1.5978 ± 0.0029 | 0.0732 ± 0.0020 | 22.59 ± 0.49 | 1.6065 ± 0.0030 | 5.29 ± 0.11 | 5.11 ± 0.14 |
| Bu4-4.3 | 17.1 | 0.06908 ± 0.00041 | 1.084 ± 0.011 | 1.6864 ± 0.0030 | 0.0904 ± 0.0029 | 18.33 ± 0.45 | 1.6981 ± 0.0031 | 6.26 ± 0.15 | 5.99 ± 0.19 |
| Bu4-4.4 | 17.7 | 0.07566 ± 0.00044 | 0.2305 ± 0.0043 | 1.5338 ± 0.0025 | 0.0912 ± 0.0020 | 92.2 ± 2.6 | 1.5439 ± 0.0025 | 6.73 ± 0.15 | 6.67 ± 0.15 |
| Bu4-2.6 | 18.8 | 0.06562 ± 0.00037 | 1.196 ± 0.013 | 1.5433 ± 0.0035 | 0.1100 ± 0.0028 | 19.16 ± 0.35 | 1.5558 ± 0.0036 | 8.38 ± 0.14 | 8.04 ± 0.21 |
| Bu4-1.3 | 19.7 | 0.07358 ± 0.00055 | 0.4260 ± 0.0062 | 1.4811 ± 0.0072 | 0.1060 ± 0.0021 | 56.7 ± 1.3 | 1.4922 ± 0.0073 | 8.19 ± 0.16 | 8.08 ± 0.17 |
| Bu4-2.7 | 20.1 | 0.07570 ± 0.00047 | 0.2362 ± 0.0059 | 1.4652 ± 0.0024 | 0.1048 ± 0.0029 | 103.3 ± 3.7 | 1.4759 ± 0.0025 | 8.13 ± 0.23 | 8.07 ± 0.23 |
| Bu4-1.4 | 20.4 | 0.06797 ± 0.00052 | 0.5555 ± 0.0055 | 1.4908 ± 0.0079 | 0.1050 ± 0.0019 | 40.00 ± 0.67 | 1.5020 ± 0.0081 | 8.10 ± 0.14 | 7.95 ± 0.16 |

[a] The ages were calculated using the half-lives of Cheng et al. (2000), and the correction for the detrital contamination assumes an average $^{232}$Th/$^{238}$U weight ratio of the upper continental crust of 3.8 ± 1.9 (Wedepohl, 1995) as well as $^{230}$Th, $^{234}$U and $^{238}$U in secular equilibrium.

### 4.2 Stable isotope and trace element analyses

#### 4.2.1    Stalagmite Bu4

The results of the $\delta^{18}$O, $\delta^{13}$C and trace element (Mg, P, Sr, and Ba) analyses of stalagmite Bu4 are shown in Figure 5. The $\delta^{18}$O and $\delta^{13}$C records of stalagmite Bu4 have an average temporal resolution of 5.3 years during the time span from 7.99 to 8.08 ka BP. The total range of the $\delta^{18}$O values during this period is between -4.6 ‰ and -6.3 ‰. The lowest $\delta^{18}$O values are reached between 8.05 and 8.02 ka BP. The total range of the $\delta^{13}$C values is between -7.5 and -9.8 ‰, and the pattern of the $\delta^{13}$C record is similar to the oxygen isotope record of the sample. Thus, the lowest $\delta^{13}$C values are also reached between 8.05 and 8.02 ka BP.

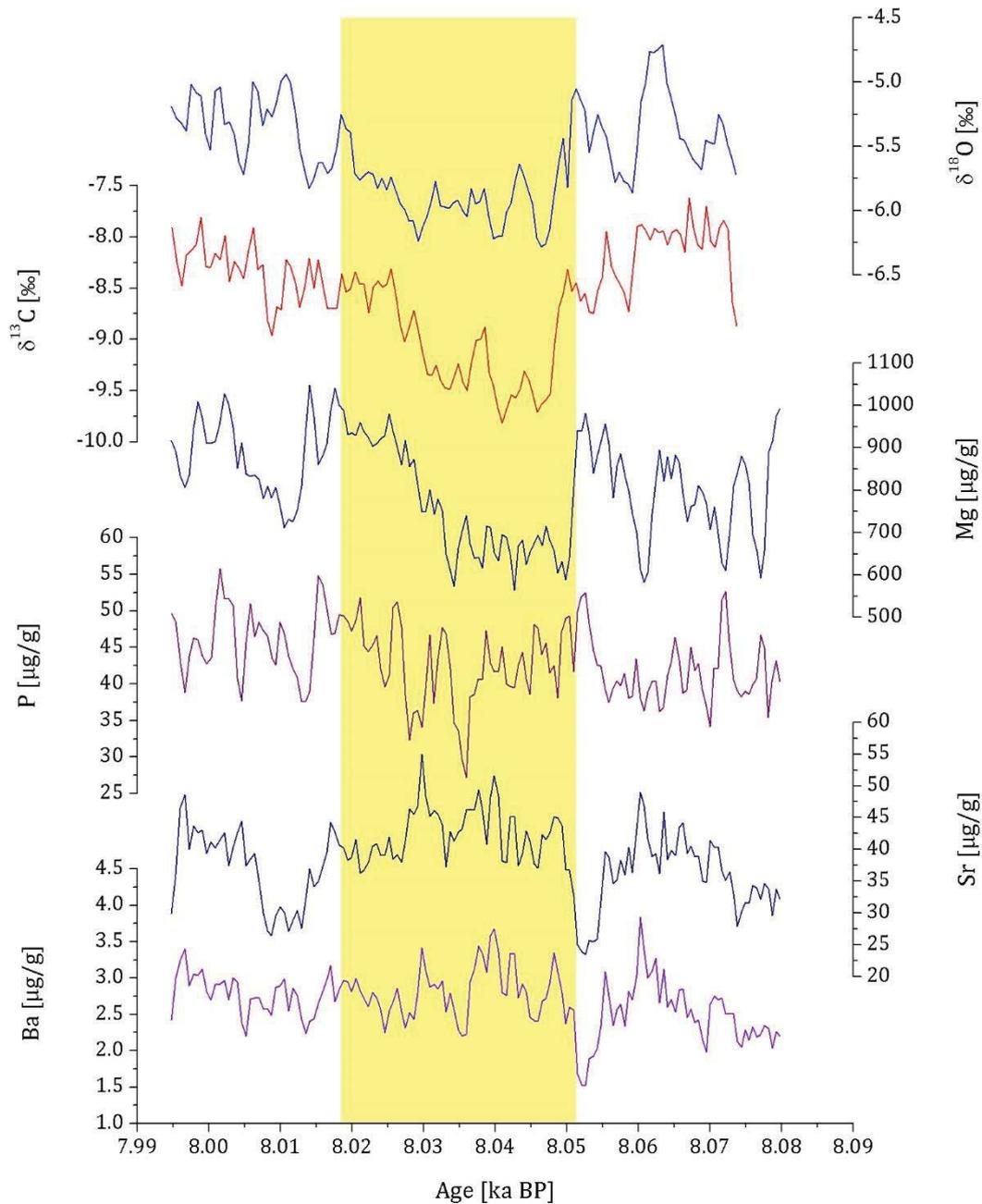

*Figure 5: Results of the stable isotope and trace element analyses (Mg, P, Sr, and Ba) of stalagmite Bu4. The trace element data were smoothed with a 10-point running median and have an average temporal resolution of 3.8 years in the period from 8.08 to 7.99 ka BP. The yellow bar highlights the 8.2 ka event recorded in the δ¹⁸O values.*

The Mg concentration of the sample varies between 500 and 1100 µg g⁻¹. Between 8.08 and 8.05 ka BP, the record shows three minima with values of approximately 600 µg g⁻¹. In the following time span from 8.05 to 8.03 ka BP, the record shows again relatively low values between 500 and 700 µg g⁻¹, similar to the δ¹⁸O and δ¹³C records. Subsequently, at 8.015 ka BP, the Mg concentration decreases again to 700 µg g$^{-1}$. All negative excursions in the Mg content have an abrupt decrease in common, followed by a more gradual increase. The decrease of the Mg content at 8.015 ka BP is also recorded in the Sr content of stalagmite Bu4. In contrast, the concentrations of P and Ba are rather stable and show no distinctive features during the time span from 8.08 to 7.99 ka BP. Exceptions are a negative peak in the Ba content between 8.055 and 8.05 ka BP, which is also recorded in the Sr content of the sample, and two negative peaks in P concentration between 8.035 and 8.03 ka BP. The pattern of the U content of stalagmite Bu4 is similar to P and is, thus, not included in Figure 5.

### 4.2.2 Stalagmite HLK2

The results of the δ$^{18}$O, δ$^{13}$C and trace element (Mg, Sr, and Ba) analyses of stalagmite HLK2 are shown in Figure 6. The stable isotope records of stalagmite HLK2 have an average temporal resolution of 3.8 years between 8.3 and 7.8 ka BP. The total range of the δ$^{18}$O values is between -5.0 ‰ and -7.0 ‰. The δ$^{18}$O record starts at 8.3 ka BP with values between -5.8 ‰ and -6.2 ‰, followed by a negative shift between 8.25 and 8.16 ka BP with a minimum of -6.56 ‰. Subsequently, the record shows a further negative excursion, which lasted for approximately 130 years. Between 7.95 and 7.75 ka BP, the δ$^{18}$O values are higher than prior to this negative excursion. However, the record also shows two short negative shifts around 7.93 and 7.84 ka BP. The total range of the δ$^{13}$C values is ‐ 8.9 to -7.0 ‰. The record starts with values of approximately ‐ 8.7 ‰, and then shows a generally positive trend towards values up to -7.2 ‰ at 7.8 ka BP.

The Mg concentration of the sample varies between 300 and 500 µg g$^{-1}$ and shows an increasing trend in the time span from 8.35 to 7.75 ka BP with constant values between 8.15 and 7.97 ka BP. The Ba content of the sample shows an opposite trend from around 5 µg g$^{-1}$ at 8.35 ka BP to 3 µg g$^{-1}$ at 8.15 ka BP. Afterwards, the Ba concentration is relatively stable with values between 2 and 3 µg g$^{-1}$. The Sr content of HLK2 varies between 15 and 20 µg g$^{-1}$ and shows more variability between 8.35 and 8.0 ka BP than between 8.0 and 7.75 ka BP. The evolution of the P and U concentration of HLK2 is relatively similar to the Ba concentration, and the lowest values are reached between 8.2 and 8.0 ka BP.

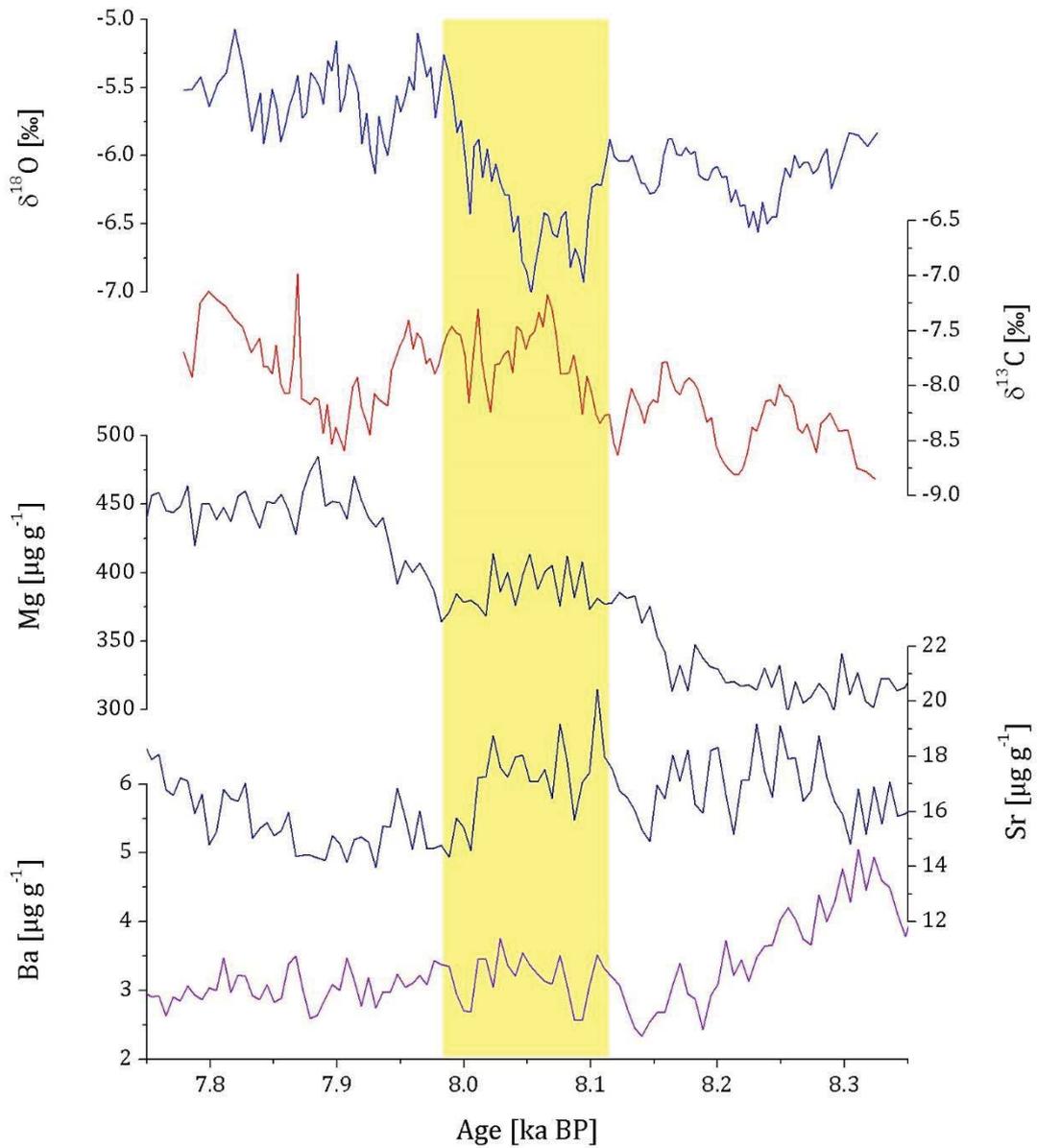

*Figure 6: Results of the stable isotope and trace element analyses (Mg, Sr, and Ba) of stalagmite HLK2. The trace element data were smoothed with a 10-point running median and have an average temporal resolution of 5.9 years during this time span. The yellow bar highlights the 8.2 ka event recorded in the δ¹⁸O values.*

### 4.2.3  Stalagmite TV1

The results of the $\delta^{18}O$, $\delta^{13}C$ and trace element (Mg, P, Sr, and Ba) analyses of stalagmite TV1 are shown in Figure 7. The $\delta^{18}O$ and $\delta^{13}C$ records of stalagmite TV1 have the same average temporal resolution as stalagmite HLK2 (3.8 years) during the time interval from 8.5 to 7.7 ka BP. The total range of the $\delta^{18}$O values during this period is between -5.3 and -7.2 ‰. The $\delta^{18}$O record shows several negative excursions in this period. It starts at 8.4 ka BP with values of about -5.6 ‰, which then decrease to ‐ 6.2 ‰ at 8.35 ka BP. Subsequently, the $\delta^{18}$O values show a distinct negative excursion for approximately 182 years with a minimum value of -7.11 ‰ at 8.17 ka BP. Between 8.09 to 7.95 ka BP, the $\delta^{18}$O values are relatively stable (-5.8 ‰ to -5.6 ‰). Afterwards, the values fluctuate again and decrease to -6.45 ‰ at 7.8 ka BP. The $\delta^{13}$C values vary between -8.5 and -9.0 ‰ and show a distinct positive peak lasting several decades with a maximum value of -7.44 ‰ at 8.25 ka BP. Subsequently, the $\delta^{13}$C values are relatively constant between -8.5 and -9.0 ‰, before they start to fluctuate, similarly to the $\delta^{18}$O values. After a decrease to -10.18 ‰ at about 7.8 ka BP, the record ends with values up to -8.2 ‰.

The Mg concentration of the sample varies between 600 and 1100 µg g$^{-1}$ between 8.45 and 7.85 ka BP. At the beginning of the record, between 8.45 and 8.2 ka BP, the values vary from 850 to 1100 µg g$^{-1}$, before the Mg concentration decreases to values of 600 to 800 µg g$^{-1}$. The Ba concentration varies between 3 and 6 µg g$^{-1}$, and shows two distinct negative excursions at ca. 8.2 ka BP (minimum concentration of 1.03 µg g$^{-1}$) and at 8.0 ka BP (minimum value of 1.56 µg g$^{-1}$). The Sr concentration (15-55 µg g$^{-1}$) shows the same two negative peaks as the Ba concentration. The evolution of the U concentration of sample TV1 is relatively similar to the P concentration and is, thus, not included in Figure 7.

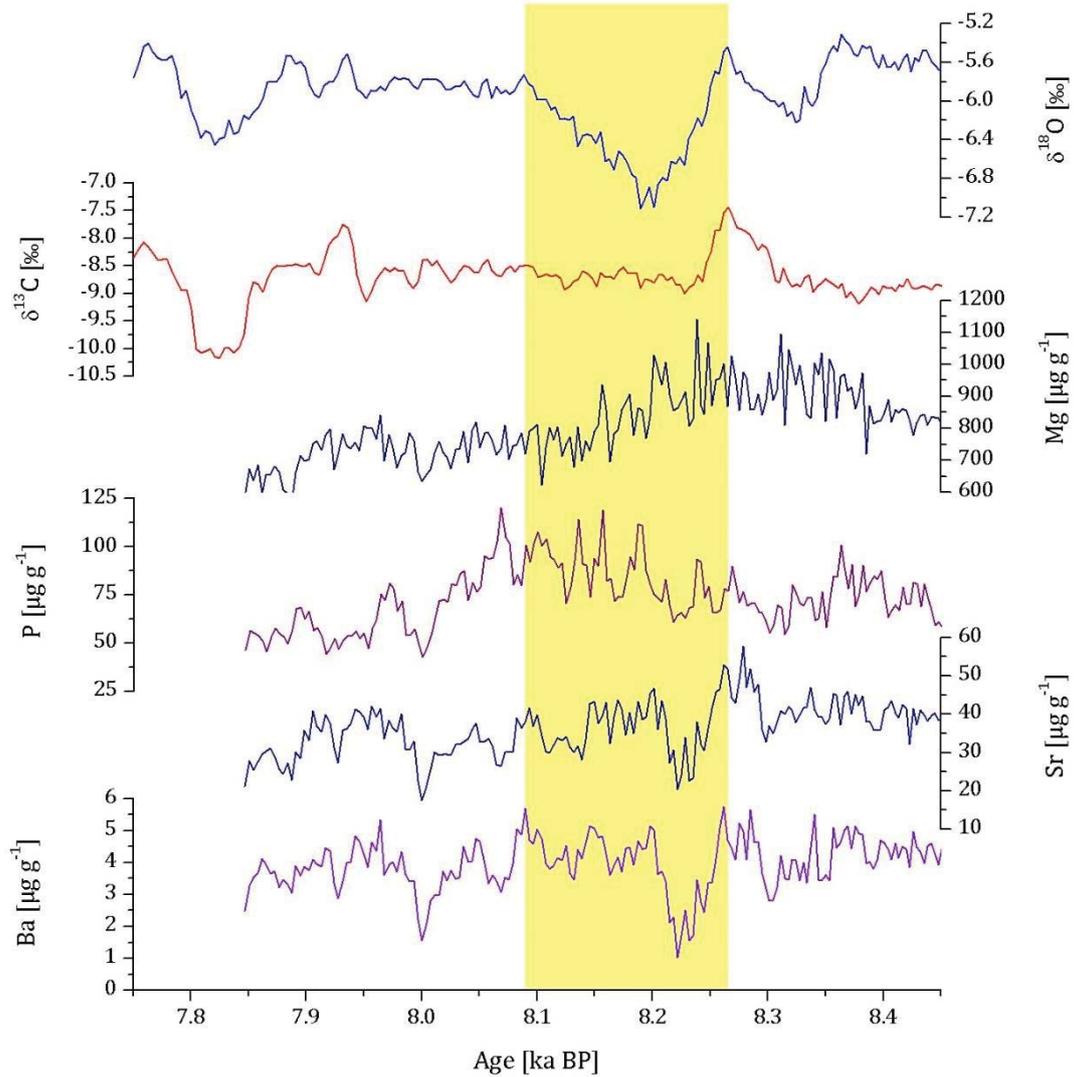

*Figure 7: Results of the stable isotope and trace element analyses (Mg, P, Sr, and Ba) of stalagmite TV1. The trace element data were smoothed with a 10-point running median and have an average temporal resolution of 3.4 years during 8.45 to 7.85 ka BP. The yellow bar highlights the 8.2 ka event recorded in the $\delta^{18}O$ values.*

4.2.4   Comparison of the high- and low-resolution data

Fig. 8 shows a comparison of the new high-resolution $\delta^{18}O$ data from this study with the lower resolution published $\delta^{18}O$ records (Fohlmeister et al., 2012; Mischel et al., 2017) on the depth scale. It is obvious that the general evolution of the low-resolution $\delta^{18}O$ values in all three records is reproduced by the new data. However, the absolute value of the negative excursion associated with the 8.2 ka event is strongly affected by the resolution. This is particularly evident for the two speleothems from the Herbstlabyrinth cave system, which were sampled at substantially lower resolution in the previous study (i.e., a factor of 10). In both HLK2 and and TV1, the minimum $\delta^{18}O$ value of the high-resolution data across the event is by ca. 0.5 ‰ lower than that of the low-resolution data (Fig. 8). Considering the temperature dependence of oxygen isotope fractionation between water and calcite (ca. -0.25 ‰/°C, Hansen et al., 2019; Tremaine et al., 2011; Kim and O'Neil, 1997), this would also have strong implications for potential temperature changes during the 8.2 ka event deduced from the high- and low-resolution $\delta^{18}O$ data. In addition, the structure of the event is – of course – much better revealed by the high-resolution data. For instance, the low-resolution $\delta^{18}O$ values of stalagmite HLK2 hardly show the 8.2 ka event as a 'whole event' and a 'central event', which is clearly visible in all three high-resolution $\delta^{18}O$ records (Fig. 8, see below for further details). This must also be kept in mind when comparing the new high-resolution $\delta^{18}O$ values with the longer, lower resolution $\delta^{18}O$ records covering large parts of the Holocene (Fig. 1).

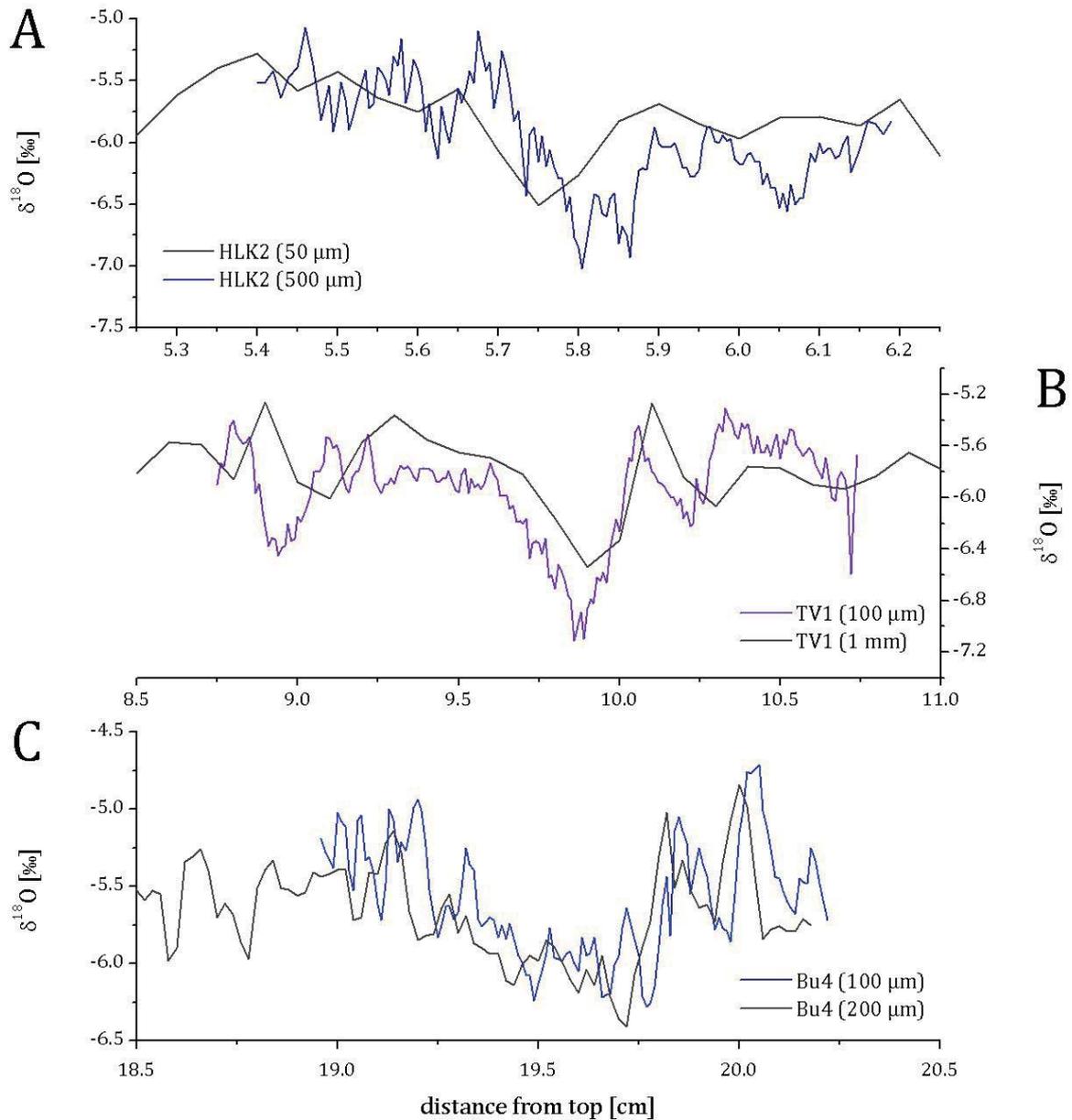

Figure 8: Comparison of the new high-resolution δ¹⁸O records over the 8.2 ka event from this study with the previously published lower resolution δ¹⁸O records (Fohlmeister et al., 2012; Mischel et al., 2017) on a depth scale. The resolution of the individual records is indicated.

## 5. Discussion

### 5.1 The expression and timing of the 8.2 ka event in the δ¹⁸O records

In all three δ¹⁸O records, the 8.2 ka event is clearly recorded as a pronounced negative excursion (Fig. 9). Interestingly, the 8.2 ka event in all three stalagmites can be divided into a 'whole event' and a 'central event', as described for Greenland ice cores (Thomas et al., 2007, see below for details). Unfortunately, although $^{230}$Th/U-dating with MC-ICP-MS generally enables a higher precision than TIMS, the age models of all three stalagmites are associated with considerable uncertainty (between 0.1 and 0.2 ka, Table 2), which is mainly due to the combination of low U content and relatively high Th contamination. This is evident from Fig. 9, where the arrows indicate the uncertainty of the timing of the minimum $\delta^{18}$O values during the 8.2 ka event based on the 95%-confidence limits of the age models. Unfortunately, this prevents conclusions about the timing and duration of the 8.2 ka cooling event. Therefore, we focus on the structure and the climate conditions of the 8.2 ka event in the following and only briefly discuss the chronological implications here.

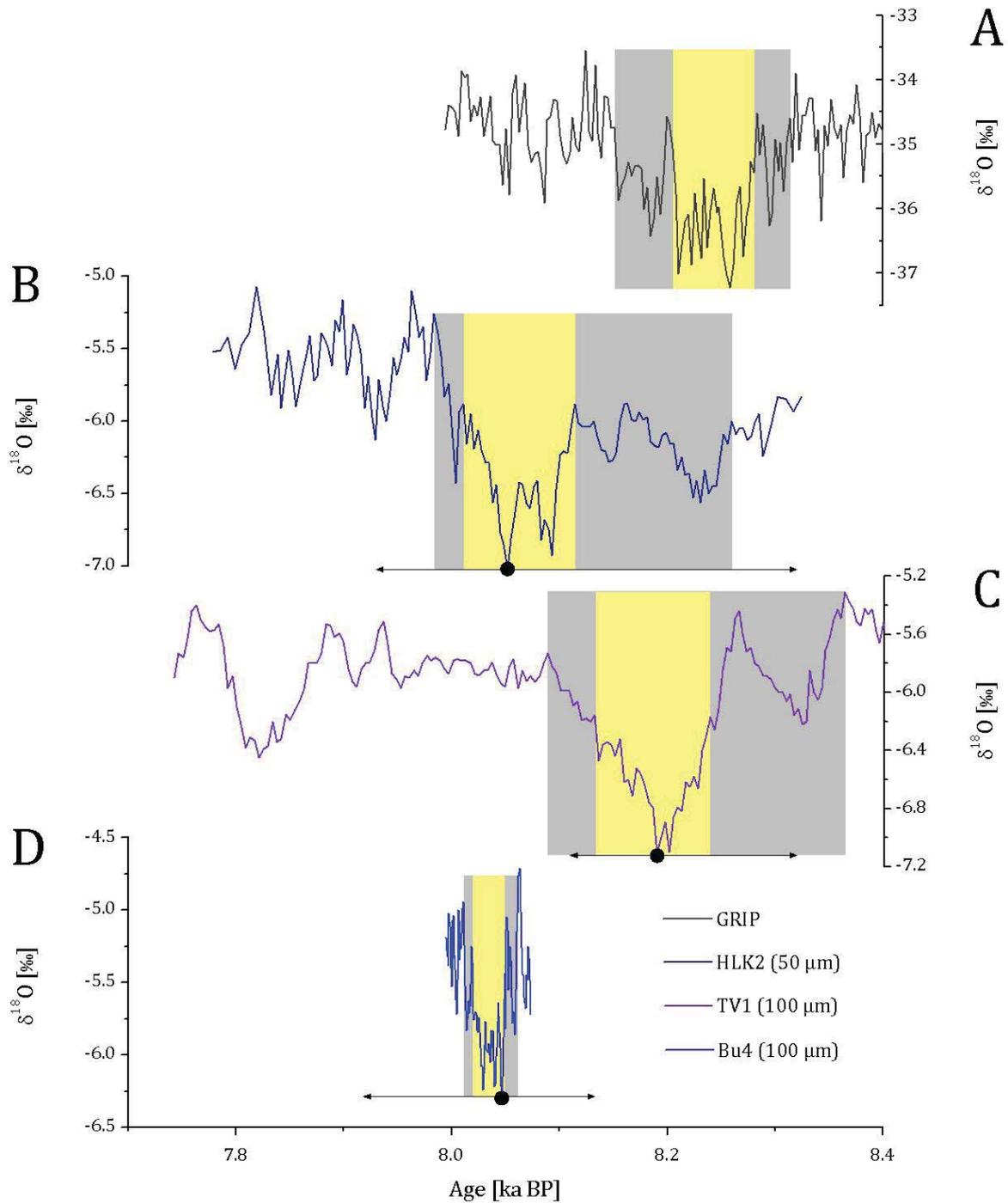

Figure 9: Comparison of the 8.2 ka cooling event in the GRIP $\delta^{18}O_{ice}$ record (A; Thomas et al., 2007) and the $\delta^{18}O$ records of the three speleothems: HLK2 (B) and TV1 (C) from the Herbstlabyrinth cave system as well as Bu4 (D) from Bunker Cave. The arrows indicate the dating uncertainties of the minimum $\delta^{18}O$ values during the 8.2 ka event. Yellow boxes indicate the timing of the 'central event' in the individual $\delta^{18}O$ records and light grey boxes mark the timing of the 'whole event'.

Table 2 shows a comparison of the timing and duration of the 'whole event' and 'central event' in the three individual stalagmites. As stated above, due to the relatively large uncertainty, conclusions about the timing and duration are difficult. Nevertheless, the timing of the event appears later in Bu4 than in the speleothems from the Herbstlabyrinth and the Greenland ice core record (8.18 ka BP at the earliest, Table 2). However, since the 8.2 ka event is contained in the relatively short bottom section of Bu4 below the hiatus at 18.5 cm (Figs. 1, 2 and 4) and considering the uncertainties of the individual ages (Fig. 4), we refrain from further interpreting this observation. The duration of the event also varies between the individual stalagmites and is much shorter in stalagmite Bu4 (50 years for the 'whole event' and 30 years for the 'central event', Table 2) than in the two speleothems from the Herbstlabyrinth cave system (180/272 years in HLK2/TV1 for the 'whole event' and 100/85 years in HLK2/TV1 for the 'central event').

Table 2: Timing, duration, mean $\delta^{18}O$ and $\delta^{13}C$ values, their standard deviation as well as the minimum and maximum $\delta^{18}O$ and $\delta^{13}C$ values of the 8.2 ka cooling event ('whole event' and 'central event') in the individual speleothems.

|  | Bu4 | HLK2 | TV1 |
|---|---|---|---|
| Start of the 'whole event' [ka] | 8.06 ± 0.12 | 8.16 ± 0.20 | 8.36 ± 0.14 |
| Start of the 'central event' [ka] | 8.05 ± 0.12 | 8.11 ± 0.20 | 8.24 ± 0.10 |
| End of the 'central event' [ka] | 8.02 ± 0.12 | 8.01 ± 0.20 | 8.155 ± 0.090 |
| End of the 'whole event' [ka | 8.01 ± 0.13 | 7.98 ± 0.19 | 8.088 ± 0.083 |
| Duration of the 'whole event' [years] | 50 | 180 | 272 |
| Duration of the 'central event' [years] | 30 | 100 | 85 |
| Mean $\delta^{18}O$ [‰, complete record] | -5.59 | -5.97 | -5.94 |
| Std. Dev. $\Delta^{18}O$ [‰, complete record] | 0.38 | 0.42 | 0.37 |
| Mean $\delta^{18}O$ [‰, 'whole event'] | -5.72 | -6.24 | -6.17 |
| Std. Dev. $\Delta^{18}O$ [‰, 'whole event'] | 0.34 | 0.38 | 0.43 |
| Mean $\delta^{18}O$ [‰, 'central event'] | -5.90 | -6.43 | -6.67 |
| Std. Dev. $\Delta^{18}O$ [‰, 'central event'] | 0.23 | 0.31 | 0.24 |
| Minimum $\delta^{18}O$ [‰] | -6.28 | -7.02 | -7.11 |
| Maximum $\delta^{18}O$ [‰] | -4.71 | -5.07 | -5.31 |
| Mean $\delta^{13}C$ [‰, complete record] | -8.61 | -8.01 | -8.70 |
| Std. Dev. $\Delta^{13}C$ [‰, complete record] | 0.55 | 0.40 | 0.47 |
| Mean $\delta^{13}C$ [‰, 'whole event'] | -8.82 | -7.87 | -8.57 |
| Std. Dev. $\Delta^{13}C$ [‰, 'whole event'] | 0.51 | 0.35 | 0.36 |
| Mean $\delta^{13}C$ [‰, 'central event'] | -9.08 | -7.79 | -8.73 |
| Std. Dev. $\Delta^{13}C$ [‰, 'central event'] | 0.46 | 0.33 | 0.11 |
| Minimum δ¹³C [‰] | -9.82 | -8.85 | -10.18 |
| Maximum δ¹³C [‰] | -7.62 | -6.98 | -6.79 |
| Resolution [years] | 5.3 | 3.8 | 3.8 |
| Time Range [ka BP] | 8.08-7.99 | 8.3-7.8 | 8.5-7.7 |

Table 2 also shows the mean and standard deviation of the δ¹⁸O and δ¹³C values of the complete record, the 'whole event' and the 'central event' are shown. The stalagmites from the Herbstlabyrinth cave system show similarities in their δ¹⁸O values. The mean δ¹⁸O values of stalagmite Bu4 are different, both for the complete record (ca. 0.35 ‰ higher) and for the 'central event' (ca. 0.77 ‰ higher). Interestingly, this difference is in the same range as today. At the Herbstlabyrinth, the mean drip water δ¹⁸O values are ca. -8.6 ‰ (Mischel et al., 2015), which is ca. 0.6 ‰ lower than the mean δ¹⁸O value of the drip water at Bunker Cave, which is ca. -8.0 ‰ (Riechelmann et al., 2011). In contrast to the mean δ¹⁸O values, the mean δ¹³C values of the complete record show a higher similarity between stalagmites Bu4 and TV1 (-8.61 ‰ and -8.70 ‰). The minimum δ¹³C value of HLK2 is -8.85 ‰ and thus about 1 ‰ higher than in the other two stalagmites.

Despite of the chronological uncertainties, the structure of the 8.2 ka event is similar between all three speleothems (Fig. 9). In all three stalagmites, the δ¹⁸O records show a short negative excursion before the central part of the event. We interpret this negative excursion as the start of the 'whole event'. Many other records also suggest the occurrence of two events. Sediment cores studied by Holmes et al. (2016) show two negative excursions in their δ¹⁸O records. The asymmetrical pattern of the event, with a rapid onset and more gradual ending, is also similar to the 'central event' in the speleothems from the two cave systems. Furthermore, Domínguez-Villar et al. (2012, 2017) show a δ¹⁸O record of a stalagmite from Kaite Cave, which also indicates two events, the first at 8.34 to 8.32 ka BP and the second between 8.21 and 8.14 ka BP (BP = AD 1950). These two events coincide with the two negative excursions in the δ¹⁸O record from stalagmite TV1 (Figure 10). In addition, Ellison et al. (2006) analysed the relative abundance of foraminifera from a North Atlantic deep-sea sediment core as well as their δ¹⁸O values and also found that the 8.2 ka event is marked by two distinct cooling events at 8490 and 8290 ka BP. In addition, the ostracod δ¹⁸O records from Lake Ammersee (Germany) and the pre-Alpine Mondsee (Austria) indicate a short negative excursion prior to the main 8.2 ka event (Figure 10; Andersen et al., 2017).

The interpretation of oxygen isotope records from speleothems is complicated by various effects that potentially influence speleothem $\delta^{18}O$ signals. These include processes occurring in the ocean, atmosphere, soil zone, epikarst as well as the cave system (Lachniet, 2009). The investigation of modern drip water and cave monitoring contribute to a better understanding of the specific processes (Baker et al., 2014; Mischel et al., 2015; Riechelmann et al., 2011, 2017). Mischel et al. (2017) conclude that the $\delta^{18}O$ values of the speleothems from the Herbstlabyrinth cave system reflect large-scale climate variability in the North Atlantic region. This is in agreement with Fohlmeister et al. (2012), who showed that the central European location of Bunker Cave, which is located only about 85 km away from the Herbstlabyrinth cave system, is as well suited for the detection of precipitation and temperature variations in relation to the variations in the North Atlantic region. In addition, Fohlmeister et al. (2012) interpreted the variability in the $\delta^{18}O$ records from the Bunker Cave stalagmites as changes in winter temperature and amount of winter precipitation, with more positive $\delta^{18}O$ values during dry and cold winters and more negative values during more humid and warmer winters.

Due to the influx of cold meltwater before the 8.2 ka event, the surface water of the North Atlantic exhibited isotopically depleted $\delta^{18}O$ values for a period of several decades (Fairchild et al., 2006a). Thus, the investigation of this cooling event requires speleothems whose $\delta^{18}O$ values are sensitive to variations in the $\delta^{18}O$ values of the North Atlantic (Fairchild et al., 2006a). This applies to the speleothems from both the Herbstlabyrinth cave system and Bunker Cave. Von Grafenstein et al. (1998, 1999) conclude that the 8.2 ka event led to a depletion of the $\delta^{18}O$ values of precipitation in central Europe, which is also recorded in the $\delta^{18}O$ record of stalagmite Bu4 from Bunker Cave (Fohlmeister et al., 2012). More negative $\delta^{18}O$ values of precipitation may result from several processes. Lower $\delta^{18}O$ values of North Atlantic sea-surface water, which is the major moisture source for Central Europe, are one possibility. Other potential processes include cooler atmospheric temperatures and persistent changes in atmospheric circulation. In the two speleothems from the Herbstlabyrinth cave system, the 8.2 ka event is only recorded in their $\delta^{18}O$ records. These negative excursions in the $\delta^{18}O$ records can also be interpreted as a result of a change in the isotopic composition of the rainfall over the Herbstlabyrinth cave system because of a change in the isotopic composition of the North Atlantic.

Figure 10 shows a comparison of the $\delta^{18}O$ records from stalagmites Bu4, HLK2 and TV1 with $\delta^{18}O$ records from other climate archives: the $\delta^{18}O$ record from the GRIP ice core in Greenland, the ostracod δ¹⁸O record from Lake Ammersee, the δ¹⁸O record from stalagmite K3 from Katerloch Cave in Austria and a composite δ¹⁸O record based on four speleothems from Kaite Cave in northern Spain. The best agreement with these records is observed for stalagmite TV1 from the Herbstlabyrinth cave system, but within the relatively large dating uncertainty, the δ¹⁸O records of Bu4 and HLK2 are also in agreement with all these records.

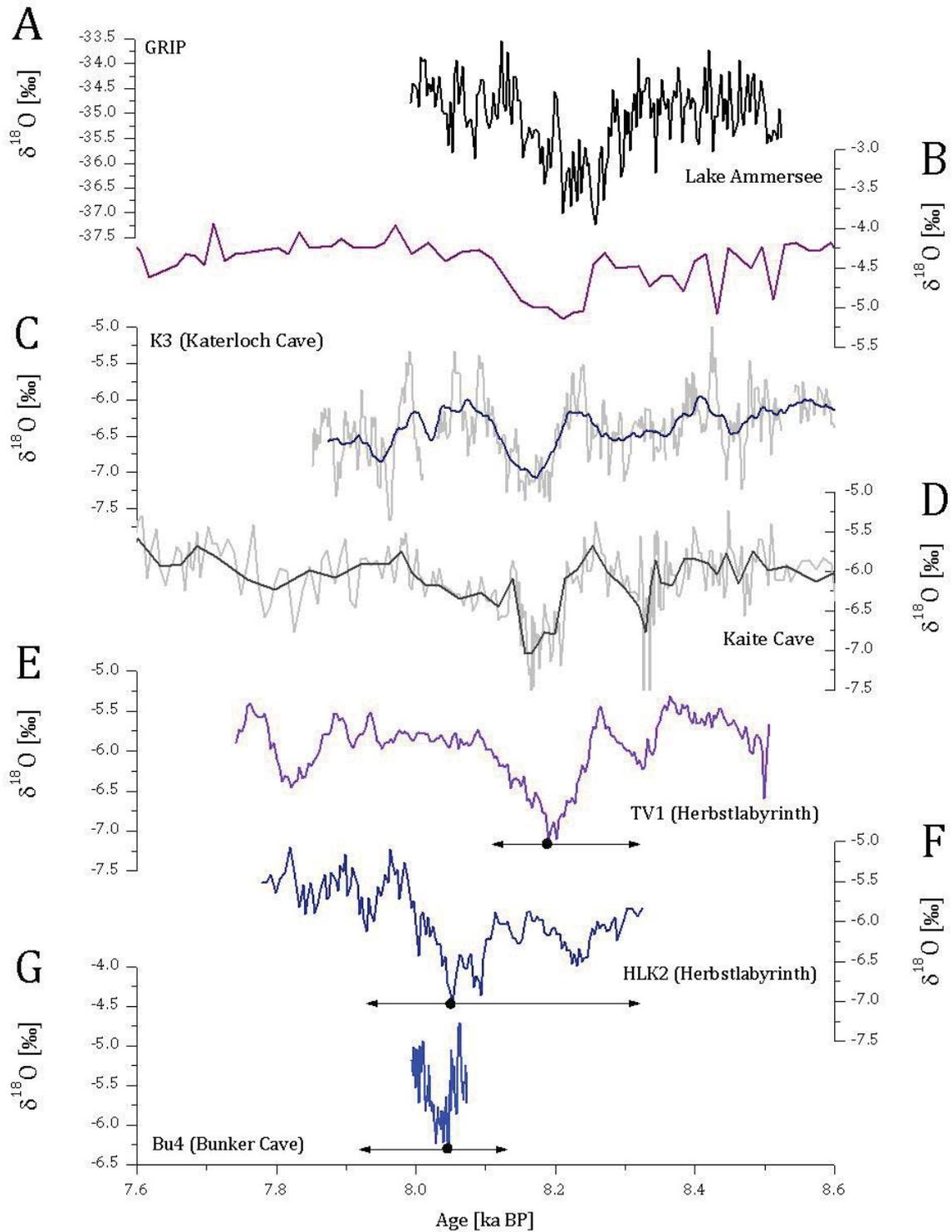

Figure 10: δ¹⁸O records of speleothems Bu4, HLK2 and TV1 compared to δ¹⁸O records from other climate archives for the time interval from 8.6 to 7.6 ka BP. A) GRIP δ¹⁸O$_{ice}$ record (Thomas et al., 2007), B) ostracod δ¹⁸O record from Lake Ammersee, southern Germany (von Grafenstein et al., 1998, 1999), C) speleothem δ¹⁸O record from Katerloch Cave, Austria (Boch et al., 2009), D) composite δ¹⁸O record based on four speleothems from Kaite Cave (Domínguez-Villar et al., 2017), E) δ¹⁸O record of stalagmite TV1 from the Herbstlabyrinth cave system, F) δ¹⁸O record of stalagmite HLK2

*from the Herbstlabyrinth cave system and G) δ¹⁸O record of speleothem Bu4 from Bunker Cave. The arrows indicate the dating uncertainties of the minimum δ¹⁸O values during the 8.2 ka event.*

## 5.2 The expression of the 8.2 ka event in the δ¹³C values

During the 8.2 ka event, only the δ¹³C values of stalagmite Bu4 show a clearly visible negative anomaly (Figure 11). The two stalagmites from the Herbstlabyrinth cave system do not show a clear peak during the event, and their pattern is also different. As δ¹⁸O values, the δ¹³C values of speleothem calcite are affected by various environmental and isotope fractionation processes. The most important factor determining the speleothem δ¹³C values is the δ¹³C value of the drip water, which is ultimately linked to the biosphere (Mischel et al., 2017). The dissolved inorganic carbon (DIC) in the drip water derives from atmospheric $CO_2$, soil $CO_2$ as well as the dissolution of the host rock (Fairchild et al., 2006b). During the latter, it is important if the dissolution occurs under conditions of an open or a closed system. In an open system, there is a continuous equilibration between the solution and an infinite reservoir of soil $CO_2$, whereas the closed system is characterised by the isolation of the solution from the soil $CO_2$ reservoir, as soon as the carbonate dissolution begins (McDermott, 2004). Under open system conditions, the δ¹³C values of the dissolved species reflects the isotopic composition of the soil $CO_2$, under closed system conditions in contrast, the isotopic composition of the host-rock influences the isotopic composition of the DIC (McDermott, 2004). Processes occurring in the soil zone (root respiration and decomposition of organic material) as well as the vegetation type (C3 or C4 plants) and productivity above the cave have also a distinct effect on the δ¹³C values (Fairchild et al., 2006a). Thus, variations in the δ¹³C values are commonly interpreted to reflect the vegetation and soil activity above the cave (McDermott, 2004; Mischel et al., 2017).

Since the two stalagmites from the Herbstlabyrinth cave system show no characteristic feature in their δ¹³C records (Figure 11), the 8.2 ka event probably had no major influence on soil activity above the cave. At first sight, this is surprising because the dramatic cooling during the event should have an effect on the vegetation above the cave, which should be reflected in soil $pCO_2$ and eventually speleothem δ¹³C values inside the cave. A recent modelling study has indeed shown that all dominant plant functional types over Europe responded to the 8.2 ka event (Li et al., 2019). However, the magnitude, timing and impact of the response is complex. In north-western Europe, for instance, the model predicts a reduction of the fraction of temperate broadleaved summergreen trees, but at the same time, a significant expansion of boreal needle-evergreen trees. In western Europe, the region of our cave sites, the fraction of temperate broadleaved summergreen trees also decreases, while the fraction of temperate broadleaved evergreen trees only shows a slight decline. This suggests that the cooler conditions during the 8.2 ka event resulted in a change of vegetation composition rather than a general reduction. Importantly, in western Europe, this change did not result in a significant change in the fractions of C3 and C4 plants (e.g., an increase in the fraction of grasses), which would be reflected in soil $pCO_2$ (and speleothem) $\delta^{13}C$ values. Thus, the effect on soil $pCO_2$ and consequently the $\delta^{13}C$ values of the speleothems at our cave sites is difficult to assess. This is supported by the compilation of European pollen records for the 8.2 ka event (Li et al., 2019), which partly show different vegetation compositions under similar climatic conditions during the 8.2 ka event (supported by the model) or – in some cases – no response at all. In summary, it is thus not unlikely that the 8.2 ka event had no major influence on soil activity at Bunker Cave and the Herbstlabyrinth cave system.

Another possibility is that the $\delta^{18}O$ values are dominated by winter precipitation, and the $\delta^{13}C$ values mainly record the vegetation activity during the vegetation period. Therefore, if the 8.2 ka cooling event was dominated by the winter season, it may have had a smaller effect on the vegetation and soil activity. The differences in the $\delta^{13}C$ records of the two stalagmites from the Herbstlabyrinth cave system also suggest that the $\delta^{13}C$ signal on the decadal to centennial time-scale is not only affected by the vegetation above the cave. The different behaviour of the $\delta^{18}O$ and $\delta^{13}C$ records is also illustrated by their low correlation coefficients of 0.17 for HLK2 and 0.27 for TV1. In contrast, speleothem Bu4 shows a negative excursion in the $\delta^{13}C$ values during the 8.2 ka event and, thus, probably indicates a different regional impact ($r_{(\delta^{18}O/\delta^{13}C)} = 0.69$) of the 8.2 event. Fohlmeister et al. (2012) assigned high $\delta^{13}C$ values in Bunker Cave speleothems to periods of low drip rates and lower vegetation density. Thus, the negative excursion in the $\delta^{13}C$ record of Bu4 suggests a well-developed vegetation and soil profile as well as higher drip rates, which should be related to more humid conditions during the 8.2 ka cooling event in Bunker Cave.

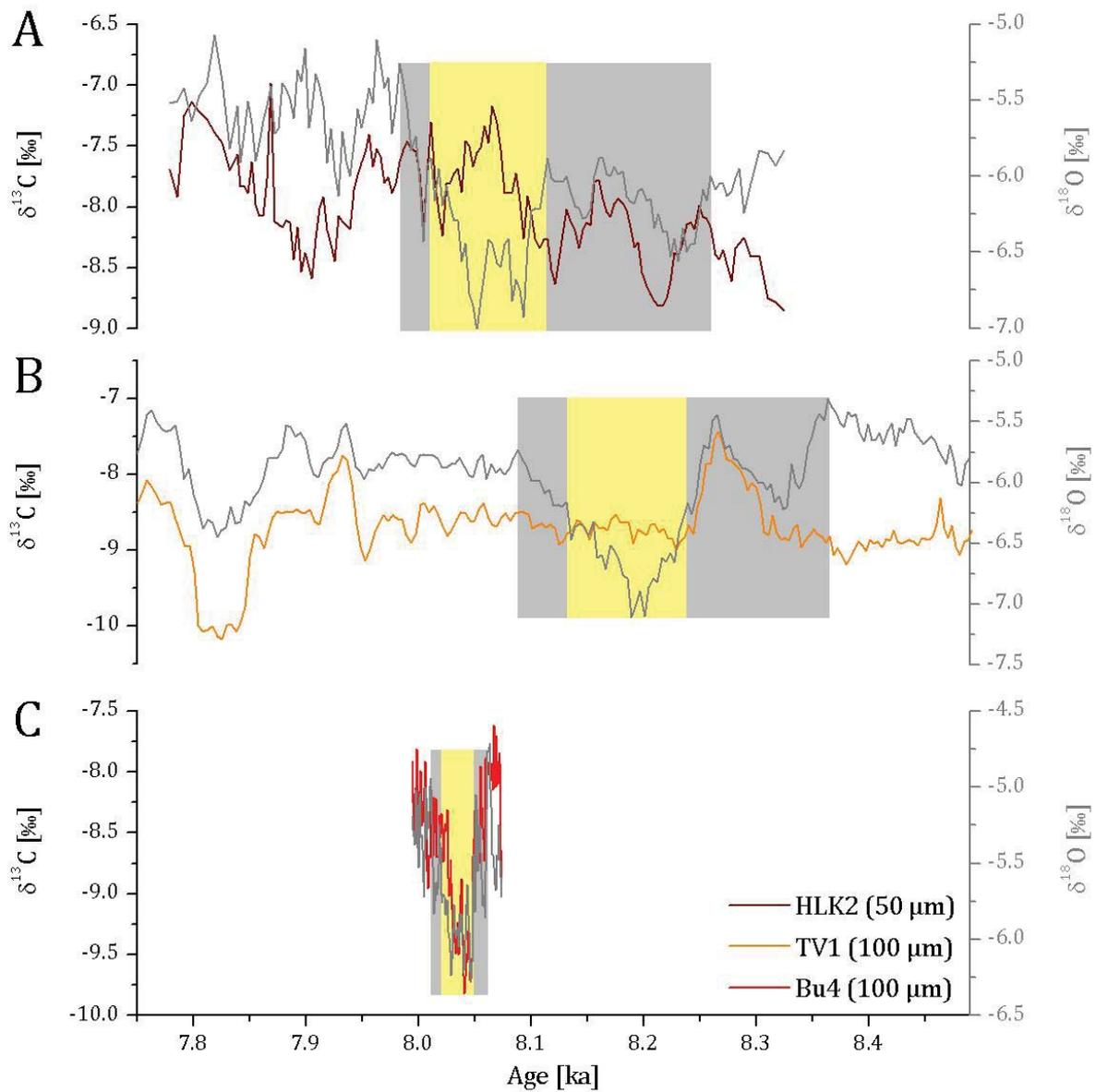

Figure 11: Comparison of the 8.2 ka cooling event in the δ¹³C records of the three speleothems: HLK2 (A) and TV1 (B) from the Herbstlabyrinth cave system as well as Bu4 (C) from Bunker Cave. The corresponding δ¹⁸O records of all three speleothems are shown in grey. The light grey boxes highlight the timing of the 'whole event' and the yellow boxes the timing of the 'central event' in the individual stalagmites.

## 5.3 The expression of the 8.2 ka event in the trace element records

Studies discussing the 8.2 ka event based on speleothem trace element data are rare. Fohlmeister et al. (2012) discussed a short negative excursion of the Mg/Ca ratios of the speleothems from Bunker Cave during the 8.2 ka cooling event. Baldini et al. (2002) reported an abrupt positive shift in Sr and a negative shift in P content of an Irish speleothem and interpreted these data as a response to a short climate anomaly with cold and dry conditions. Only some of our trace element data show distinctive features during the 8.2 ka cooling event. However, the patterns are different for the individual stalagmites. In the trace element data of stalagmite Bu4, only Mg content shows a negative excursion during the 8.2 ka event, which is correlated with the $\delta^{18}O$ and $\delta^{13}C$ values and confirms the results of Fohlmeister et al. (2012). Previous studies suggested that the Mg content of speleothems increases during prior calcite precipitation (PCP), accompanied by an increase in Sr content and in $\delta^{13}C$ (Fairchild and Treble, 2009). In addition, the Mg content of speleothems may be a useful proxy for effective precipitation because it reflects the changes in residence time in the karst aquifer (Fairchild and Treble, 2009). During dry times, when the residence time is longer due to reduced effective precipitation, the prolonged contact of the water with the host rock leads to an increased Mg content in the drip water and, thus, in speleothem calcite (Fairchild and Treble, 2009; Hellstrom and McCulloch, 2000; McDonald et al., 2004). For Bunker Cave, this has been confirmed by cave monitoring, which shows that the stalagmite Mg/Ca ratio is influenced by PCP, which in turn affects the Mg/Ca ratio of the drip water and speleothem calcite (Fohlmeister et al., 2012; Riechelmann et al., 2011). The Mg/Ca minimum during the 8.2 ka event indicates more humid conditions in the region of Bunker Cave, which is consistent with the study of Fohlmeister et al. (2012), who interpreted the short-term variations in the Mg/Ca ratios as infiltration variability above the cave. The more humid conditions during the 8.2 ka event in Bunker Cave also agree with Flohr et al. (2016), who suggest increased wetness during the 8.2 ka event north of 42° N, while aridity increased south of 42° N. Abrantes et al. (2012) also described dry conditions in the central and eastern Mediterranean regions during the event and an increase in precipitation north of 42°N, which was detected in pollen sequences. However, Magny et al. (2003b) as well as Berger and Guilaine (2009) concluded that only the mid-latitudes between about 42° and 50° N underwent more humid conditions during the 8.2 ka event, whereas the climate in northern and southern Europe became drier. Bunker Cave and the Herbstlabyrinth cave system are located slightly north of 50°N and are therefore situated in a transition zone.

The P content of Bu4, which is generally interpreted as a proxy for soil activity and wetness and usually increases during wet times because of a more productive vegetation cover (Treble et al., 2003; Mischel et al., 2017), shows no distinctive features during the 8.2 ka event and, thus, no further evidence for more enhanced humidity. The Sr and Ba contents of speleothem Bu4 are highly correlated (r = 0.72) suggesting similar processes influencing their concentration changes, but show no significant anomalies during the 8.2 ka event (Figure 5). Positively correlated Sr and Ba concentrations in speleothems have been interpreted as reflecting changes in density and productivity of the vegetation cover above the cave with increasing values reflecting wetter conditions (Hellstrom and McCulloch, 2000; Desmarchelier et al., 2006). However, as for the P concentration of speleothem Bu4, the evolution of the Sr and Ba records give no further evidence for a more productive vegetation cover above Bunker Cave or an increase in humidity during the 8.2. ka event. In contrast, Treble et al. (2003) concluded the high correlation between the Ba and Sr concentrations indicates that the annual Sr cycle is affected by the growth rate of the speleothem.

Mischel et al. (2017) interpreted the P, Ba, and U content of speleothems from the Herbstlabyrinth cave system as proxies for vegetation productivity above the cave system with higher concentrations during phases with a more productive vegetation. Higher P, Ba and U concentrations in the speleothems from the Herbstlabyrinth cave system generally coincide with lower Mg and $\delta^{13}C$ values reflecting a higher vegetation productivity and more precipitation (Mischel et al., 2017). If the climate during the 8.2 ka cooling event at the Herbstlabyrinth cave system had been more humid as described for Bunker Cave, we would expect an increase in P, Ba and U during the event and a decrease in Mg and the $\delta^{13}C$ values of stalagmites HLK2 and TV1. However, both stalagmites neither show an increase in P, Ba and U nor a decrease in Mg and $\delta^{13}C$ values. Stalagmite TV1 shows a short negative excursion in Sr and Ba, which starts at the same time as the negative excursion in the $\delta^{18}O$ values and is also evident in the trace element records of speleothem Bu4. However, both the mid-point (~ 8.22 ka BP) and the end (~ 8.20 ka BP) of the negative excursion in the Sr and Ba content are much earlier than in the $\delta^{18}O$ record. Thus, it is not clear if this peak corresponds to the 8.2 ka cooling event. The structure of the $\delta^{18}O$ record is also different than for the Ba and Sr records. In the $\delta^{18}O$ record of TV1, the 8.2 ka event has a slightly asymmetrical structure with a rapid onset and a more gradual ending. The opposite pattern is observed in the Sr and Ba content (Figure 7). Following the interpretation of Mischel et al.

(2017), the negative excursion in the Ba content of stalagmite TV1 during the 8.2 ka event probably suggests drier conditions because of a lower vegetation activity. However, the Ba content of a speleothem is not only a proxy for the density and productivity of the vegetation cover above the cave, it may also indicate changes in the growth rate of the stalagmite (Treble et al., 2003). The Sr content of a speleothem is controlled by hydrological processes and is sensitive to phases with a longer residence time of the groundwater (Treble et al., 2003). In addition, the Sr concentration increases during phases with higher growth rates (Fairchild and Treble, 2009). The highly correlated Sr content of sample TV1 ($r_{(Sr/Ba)} = 0.74$) shows the same negative peak and suggests that the Sr and Ba contents are controlled by the same environmental parameters. This strong correlation between Sr und Ba is also observed in stalagmite Bu4 from Bunker Cave, but not in the other stalagmite HLK2 from the Herbstlabyrinth cave system.

In summary, stalagmite Bu4 from Bunker Cave shows a distinct negative period in the Mg concentration, and stalagmite TV1 from the Herbstlabyrinth cave system shows a short negative excursion in the Sr and Ba content during the 8.2 ka cooling event. In contrast, stalagmite HLK2 from the same cave system shows no distinctive features in the trace element data. This different behaviour of the trace elements in the three stalagmites during the event is also evident in the principal component analysis (PCA), which was applied to the trace element and stable isotope data of all three speleothems (supplementary material). Thus, there are no obvious similarities in the trace element data of the individual stalagmites, neither within the same cave nor for the speleothems from different cave systems. These differences can be attributed to site-specific effects, such as processes occurring in the karst aquifer or the cave system, which may have a strong influence on the geochemistry of the speleothems.

### 5.4 Climate during the 8.2 ka event at the two cave sites in Central Europe

It is widely accepted that climate during the 8.2 ka event was generally cooler, drier and potentially windier (Alley and Ágústsdóttir, 2005; Mayewski et al., 2004). Climate in the mid and high latitudes of Europe was characterised by a distinct cooling, whereas climate in the low latitudes was marked by drier conditions (Mayewski et al., 2004). In addition to this strong cooling, Alley and Ágústsdóttir (2005) also describe hydrological changes in Europe, especially during winter months. The decrease in annual air temperature in Europe was around -1 to -3 °C. For instance, cooler temperatures during the 8.2 ka event are reflected in several pollen records from Europe, which show a decline in the thermophilic deciduous tree species *Corylus* (hazel) and *Quercus* (oak), as well as an increase in the cold-resistant taxa *Betula* (birch) and *Pinus* (pine; Ghilardi and O'Connell, 2013; Hede et al., 2010; Seppä et al., 2005). Von Grafenstein et al. (1998) suggested a decrease of the average annual air temperature of about -1.7 °C based on an ostracod $\delta^{18}O$ record from Lake Ammersee. Based on $\delta^{18}O$ records of two speleothems from Katerloch Cave, Austria, a temperature decrease of about -3 °C was reconstructed (Boch et al., 2009).

Based on the low-resolution Bu4 $\delta^{18}O$ record, Fohlmeister et al. (2012) estimated the temperature change during the 8.2 ka event. This was based on the decrease in the $\delta^{18}O$ values of Bu4 during the event (- 0.4 ‰), the estimated change in the $\delta^{18}O$ values of precipitation (-0.7 ‰, von Grafenstein et al., 1998) and the temperature dependence of oxygen isotope fractionation between water and speleothem calcite (ca. -0.25 ‰/°C, Mühlinghaus et al., 2009). The lower amplitude of the Bu4 $\delta^{18}O$ values compared to the $\delta^{18}O$ values of precipitation can be accounted by a temperature decrease of ca. 1.2 °C (0.3 ‰/0.25 ‰/°C = 1.2 °C). Our new high-resolution $\delta^{18}O$ data show a stronger decrease during the 8.2 ka event in all three speleothems (-0.70 ‰, Bu4; 1.05 ‰, HLK2; -1.17 ‰, TV1, Fig. 8). For the determination of the $\delta^{18}O$ amplitudes, the complete high-resolution records were used. Using the approach of Fohlmeister et al. (2012), would result in a temperature change between 0 and +4.5 °C and would, thus, rather suggest a warming during the 8.2 ka event. This strongly suggests that the change of -0.7 ‰ for the $\delta^{18}O$ values of precipitation during the 8.2 ka event by von Grafenstein et al. (1998), which is based on the Lake Ammersee record, is not valid for Bunker Cave and the Herbstlabyrinth and indicates a stronger decrease in the $\delta^{18}O$ values of precipitation for the cave regions.

The climate model simulations performed by Holmes et al. (2016) suggest a decrease in the $\delta^{18}O$ values of precipitation between -0.5 and -1.0 ‰ for the 8.2 ka event at the two cave sites. In addition, their simulations suggest a relatively low surface temperature change between +0.5 and - 0.5 °C. These amplitudes are similar to those observed in our speleothem records suggesting that the observed changes in speleothem calcite $\delta^{18}O$ values during the 8.2 ka event are mainly due to the change in the δ¹⁸O values of precipitation and only contain a minor temperature component. This implies that the variation in the δ¹⁸O values during the prominent 8.2 ka event can rather be attributed to changes in the North Atlantic than temperature changes at the cave site.

This observation may also explain why in our speleothem records, the event is mainly reflected in the δ¹⁸O values. In the speleothems from the Herbstlabyrinth cave system, the other proxies do not show a response, except from a short excursion in Sr and Ba in stalagmite TV1. In Bu4, the negative excursion in the δ¹⁸O values is accompanied by more negative δ¹³C values and a lower Mg content, which can be interpreted as more humid winter conditions. In general, however, the majority of speleothem climate proxies at the two cave sites do not suggest strong climate change during the 8.2 ka event. This either means that these proxies are not sensitive to this centennial-scale cooling event, that the climate impact of the event was lower than previously assumed or that the climate response to the event was regionally heterogenous. The latter is suggested by the different evolution of the proxies at Bunker Cave and the Herbstlabyrinth cave system, which are only 85 km apart. However, the fact that the 8.2 ka event has been detected in various records from all parts of the world, is in conflict with this hypothesis. In addition, it is questionable that such a widespread, persistent and strong climate event would not affect the majority of speleothem proxies, although the discussion of the potentially low impact of changes in vegetation composition on both soil pCO₂ and speleothem δ¹³C values shows that this may indeed be the case for some proxies. In this context, it is also interesting that the long-term δ¹⁸O records of all three stalagmites show several negative excursions of similar or even larger magnitude than the 8.2 ka event (Fig. 1). This is particularly remarkable because the 8.2 ka event is by far the largest Holocene cooling event in the Greenland ice core δ¹⁸O record (Fig. 1). As shown in section 4.2.4, sampling at higher resolution would probably even increase the magnitude of these negative excursions (Fig. 8), in particular for the speleothems from the Herbstlabyrinth, which have only been analysed at relatively low resolution in the sections younger than 7 ka (Fig. 1). This may suggest that the impact of the 8.2 ka event in Central Europe was less severe than in Greenland (in comparison to the rest of the Holocene). However, it may also be evidence for a lower (regional) impact of the 8.2 ka event at our cave sites.

Another explanation for the observation that the event is mainly reflected in the δ¹⁸O values is that the speleothem δ¹⁸O values recorded in the three stalagmites do not primarily reflect climate change at the cave site, but rather large-scale changes in the region of the moisture source (i.e., the North Atlantic). A similar explanation has been proposed to explain the changes in Chinese speleothem $\delta^{18}O$ values, which are commonly interpreted as a proxy for East Asian Monsoon strength (Pausata et al., 2011). This would also explain that the 8.2 ka event is mainly reflected in $\delta^{18}O$ records (ice cores, speleothems, ostracods). A fifth explanation is related to the seasonality of the 8.2 ka event. If the event mainly was a winter phenomenon, those proxies that are mainly affected by climate conditions during the vegetation period ($\delta^{13}C$ values as well as P, Ba, and U at the Herbstlabyrinth cave system, Mischel et al., 2017) would probably not record the event. Thus, our results can also be interpreted as suggesting a strong seasonality of the 8.2 ka event.

The comparison with the climate modelling data from Holmes et al. (2016) suggests that the majority of the change in speleothem $\delta^{18}O$ values is due to changes in the $\delta^{18}O$ values of precipitation and that temperature only played a minor role. However, based on our records, it is not possible to completely resolve which of the discussed hypotheses for the climate of the 8.2 ka event (low sensitivity of most proxies, generally lower impact, strong regional variability, changes in the region of the moisture source and strong seasonality) is most appropriate. In any case, our multi-proxy dataset shows that climate evolution during the event was probably complex, although all $\delta^{18}O$ records show a clear negative anomaly.

## 6. Conclusions

We present high-resolution multi-proxy data for the 8.2 ka event from three stalagmites (Bu4, HLK2 and TV1) from two different cave systems (Bunker Cave and Herbstlabyrinth cave system). The 8.2 ka event is clearly recorded in the $\delta^{18}O$ records of all speleothems as a pronounced negative excursion and can be defined as a 'whole event' and a 'central event'. All stalagmites show a similar structure of the event with a short negative excursion before the main 8.2 ka cooling event.

Whereas stalagmite Bu4 from Bunker Cave also shows a negative anomaly in the $\delta^{13}C$ values and Mg content during the event, the two speleothems from the Herbstlabyrinth cave system do not show distinct peaks in the other proxies, except from minor peaks in Sr and Ba in stalagmite TV1. The negative anomaly in the $\delta^{13}C$ values and Mg content of speleothem Bu4 indicates a higher vegetation productivity and higher drip rates, and, thus, suggests more humid conditions during the event at Bunker Cave. In general, however, climate change during the 8.2 ka event was not recorded by the majority of speleothem climate proxies at the two cave sites.

Comparison with climate modelling data suggests that it is likely that the 8.2 ka event had no major influence on soil activity at Bunker Cave and the Herbstlabyrinth cave system, which may explain the observed inconsistent response in speleothem $\delta^{13}C$ values. In addition, the change in speleo-them $\delta^{18}O$ values during the 8.2 ka event is mainly related to the corresponding change in the $\delta^{18}O$ values of precipitation above the cave. This implies that temperature only played a minor role.

### Acknowledgements

S. Waltgenbach and D. Scholz are thankful to the DFG for funding (SCHO 1274/9-1, 10-1, and 11-1). In addition, we thank Beate Schwager, Brigitte Stoll and Ulrike Weis for assistance in the laboratory. This project is TiPES contribution #39: This project has received funding from the European Union's Horizon 2020 research and innovation programme under grant agreement No 820970. Constructive comments of two anonymous reviewers and the Editor, A. Haywood, were very helpful to improve the manuscript.

## Appendix

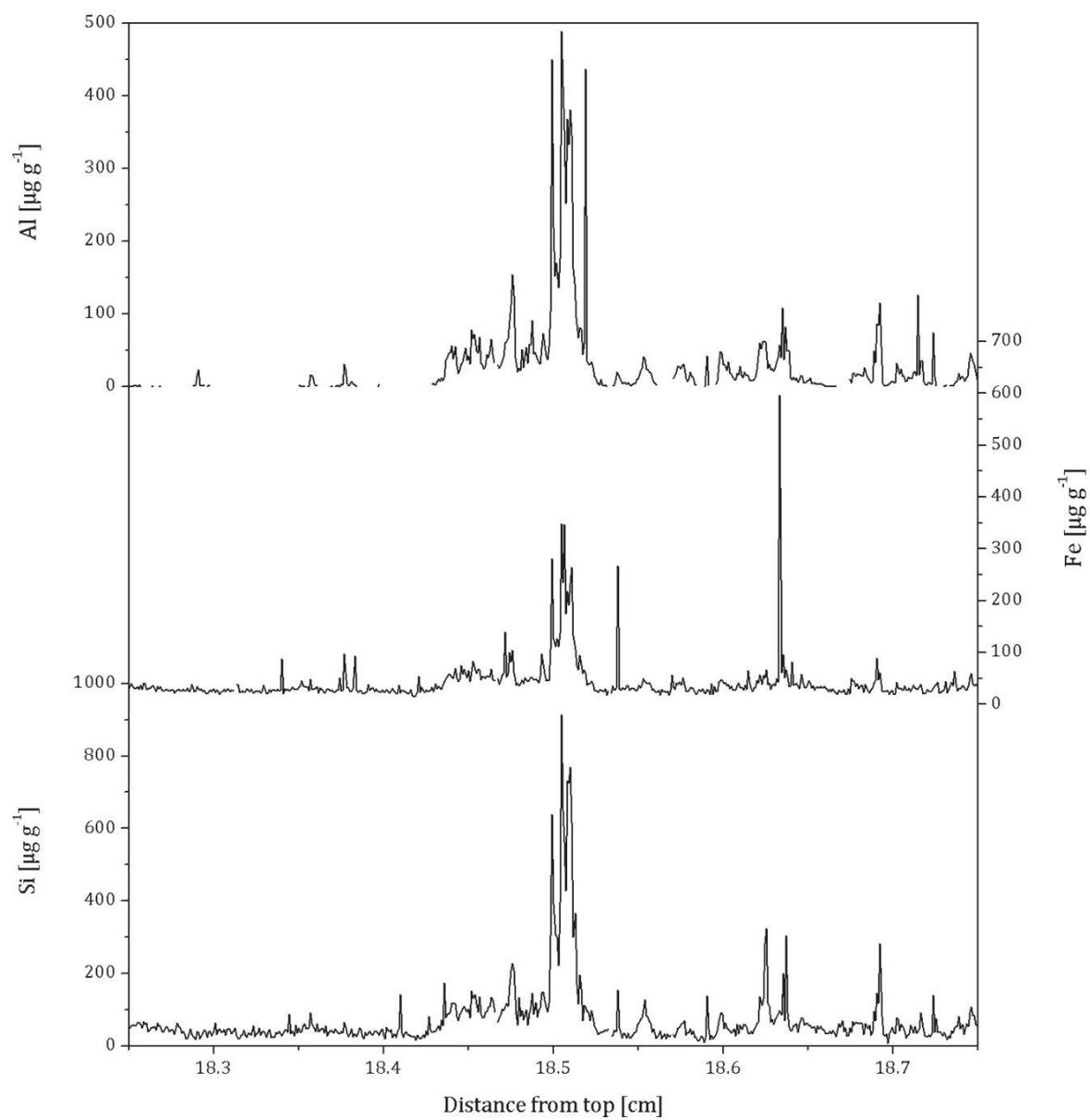

*Figure A.1: Trace element concentrations (Al, Fe and Si) at 18.5 cm distance from top indicating a potential hiatus in stalagmite Bu4 from Bunker Cave.*

### The principal component analysis (PCA)

A principal component analysis (PCA) was applied to the trace element and stable isotope data of all three speleothems to determine the main environmental processes that control the speleothem behaviour during the 8.2 ka cooling event in the individual stalagmites. For the PCA, the trace element data, which was smoothed with a 10-point running median, and the stable isotope values were normalized. In addition, the depth scale was used to adapt the resolution of the trace element and stable isotope data. The PCA was performed using the software *PAST* (**PA**leontological **ST**atistics: http://folk.uio.no/ohammer/past/; 28.03.2019), and the results of the PCA are shown in Figure A.2.

In Figure S.2-A, the highly correlated variation of $\delta^{18}O$, $\delta^{13}C$ and Mg in stalagmite Bu4 from Bunker Cave is illustrated by their clustering with moderate loadings on the first principal component (PC 1: 39 %) and low loadings on the second principal component (PC 2: 26 %). Furthermore, Sr and Ba have a moderate negative loading on PC 1 and a low to moderate loading on PC 2. Their clustering in Figure 11 reveals the high correlation (r = 0.72) of both trace elements, which was already described in Chapter 5.2 and indicates that Sr and Ba are controlled by the same environmental parameters. U and P have the highest loadings on PC 2, but only low (U: 0.023) and moderate (P: 0.32) loadings on PC 1.

In Figure S.2-B, the stable isotopes as well as Mg have the highest loadings on PC 1 (41 %), which is similar to the PCA of Bu4, but different loadings on PC 2 (30 %). $\delta^{18}O$ has a moderate (0.50), Mg a low (0.011) and $\delta^{13}C$ a low negative loading (-0.17) on PC 2. The highly correlated variation of the trace elements P, Ba and U in speleothem HLK2 from the Herbstlabyrinth cave system (r (U and Ba) = 0.85; r (P and Ba) = 0.87; r (U and P) = 0.96) is illustrated by their clustering with low to moderate negative loadings on PC 1 and moderate loadings on PC 2. Sr has a moderate negative loading on both PC 1 (-0.42) and PC 2 (-0.46). The highly correlated relationships between P, Ba and U are also observed in the PCA of the whole Holocene section of stalagmite HLK2 (Mischel et al., 2017). As already mentioned, they interpreted the P, Ba and U content as proxies for the vegetation productivity above the cave system and, thus, reflecting changes in vegetation density. Higher values of P, Ba and U in combination with lower Mg and $\delta^{13}C$ values are interpreted as reflecting a higher vegetation productivity and more precipitation (Mischel et al., 2016). Thus, the low values of P, Ba and U and the relatively high values of Mg and $\delta^{13}C$ during the 8.2 ka cooling event in stalagmite HLK2 would probably suggest dry conditions at the Herbstlabyrinth cave system during this time phase because of a lower vegetation activity above the cave.

The PCA of the other stalagmite from the Herbstlabyrinth cave system, speleothem TV1, shows different results (Figure S.2-C). The stable isotopes as well as all trace elements have positive loadings on PC 1, which explains only 34 % of the total variance, PC 2 explains 21 %. Sr and Ba have the highest loadings on the first principal component and relatively low loadings on the second principal component. In contrast, $\delta^{18}O$ and $\delta^{13}C$ have the highest loadings on PC 2 and only low loadings on PC 1. Mg, P and U have all low to moderate loadings on both PC 1 and PC 2.

The comparison of the trace element records of stalagmite Bu4 from Bunker Cave and HLK2 and TV1 from the Herbstlabyrinth cave system already showed the differences between the trace element data of the individual speleothems. The PCA further confirms these distinct differences.

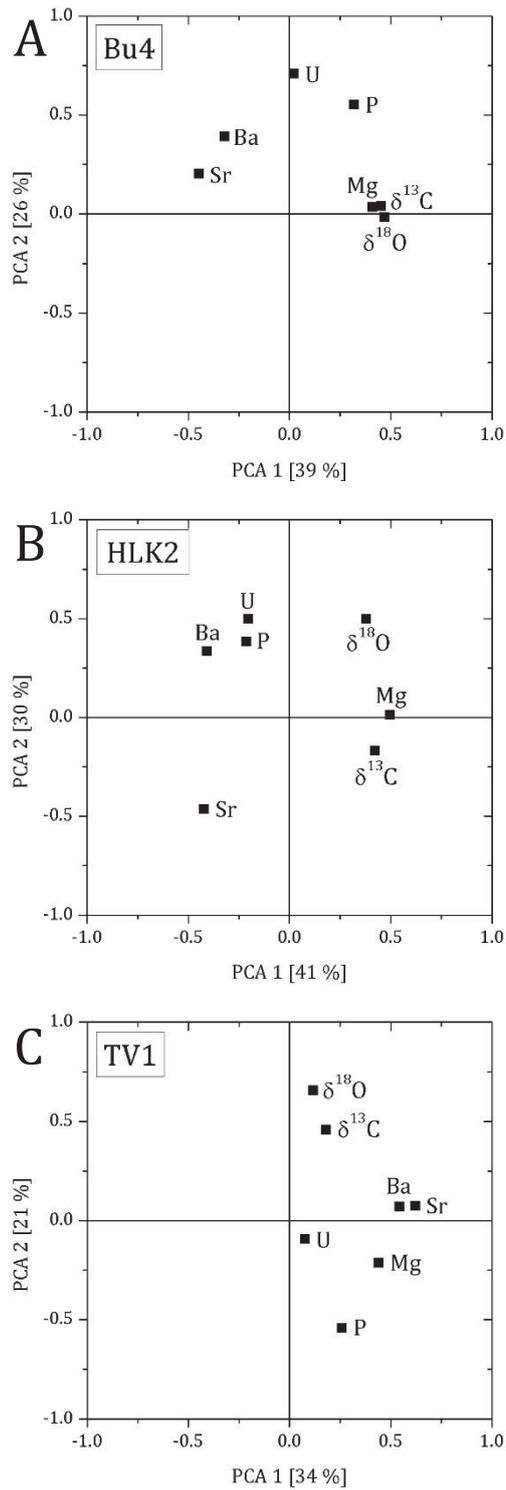

*Figure A.2: Results of the principal component analysis of stalagmites Bu4 (A) from Bunker Cave and HLK2 (B) and TV1 (C) from the Herbstlabyrinth cave system. PC 1 explains in the stalagmites between 34 and 41 % of the total variance, whereby PC 2 explains only 21 to 30 %. Only PC 1 and PC 2 are shown in the Figure and described in the supplementary material.*